\documentclass[prd,nofootinbib,english]{revtex4}

\usepackage[letterpaper,top=2cm,bottom=2cm,left=2.1cm,right=2.1cm,marginparwidth=1.75cm]{geometry}

\usepackage{graphicx,float}
\usepackage{amsmath,amssymb,amsfonts}
\usepackage{upgreek}
\usepackage{mathrsfs}
\usepackage{epsfig,color}
\usepackage[thinlines]{easytable}
\usepackage{pdfpages}
\usepackage{array}
\usepackage{cancel}
\usepackage{mathtools}
\usepackage{accents}
\usepackage{subfigure}
\usepackage{enumitem}
\usepackage[dvipsnames]{xcolor}
\usepackage{refstyle}
\usepackage{hyperref}

\hypersetup{
    colorlinks=true,
    linkcolor=blue,
    filecolor=magenta,
   citecolor=blue
}

\newcommand{\reff}[1]{(\ref{#1})}

\def\be{\begin{equation}}
\def\ee{\end{equation}}
\def\bea{\begin{eqnarray}}
\def\eea{\end{eqnarray}}

\def\mH{\mathcal{H}}

\def\mA{\mathcal{A}}

\def\mU{\mathcal{U}}
\def\mV{\mathcal{V}}

\def\zi{\mathrm{i}}
\def\ze{\mathrm{e}}

\def\zd{\mathrm{d}}
\def\zf{\mathrm{f}}
\def\zs{\mathrm{s}}

\def\zA{\mathrm{A}}

\def\zT{\mathrm{T}}
\def\zL{\mathrm{L}}
\def\zR{\mathrm{R}}

\def\vi{\mathfrak{i}}

\def\va{\mathfrak{a}}
\def\vb{\mathfrak{b}}
\def\vc{\mathfrak{c}}
\def\vd{\mathfrak{d}}
\def\vA{\mathfrak{A}}
\def\vB{\mathfrak{B}}
\def\vC{\mathfrak{C}}

\def\vp{\mathfrak{p}}

\def\ce{\varepsilon}
\def\sech{\mathrm{sech}}
\def\csch{\mathrm{csch}}
\def\arccosh{\mathrm{arccosh}}

\def\pA{\mathfrak{p}_{\mathrm{A}}}

\begin{document}
\hfill  USTC-ICTS/PCFT-22-27
\title{Irregular universe in the Nieh-Yan modified teleparallel gravity}

\author{Mingzhe Li}
\affiliation{Interdisciplinary Center for Theoretical Study, University of Science and Technology of China, Hefei, Anhui 230026, China}
\affiliation{Peng Huanwu Center for Fundamental Theory, University of Science and Technology of China, Hefei, Anhui 230026, China}

\author{Haomin Rao}
\affiliation{School of Fundamental Physics and Mathematical Sciences, Hangzhou Institute for Advanced Study, UCAS, Hangzhou 310024, China}
\affiliation{University of Chinese Academy of Sciences, 100190 Beijing, China}


\begin{abstract}
The Nieh-Yan modified teleparallel gravity is a model
which modifies the general relativity equivalent teleparallel gravity by a coupling between the Nieh-Yan density and an axion-like field.
This model predicts parity violations in the gravitational waves if the axion-like field has a non-trivial background,
and more importantly it is ghost free and avoids the pathologies presented in other parity-violating gravity models.
The cosmological dynamics and perturbations of the Nieh-Yan modified teleparallel gravity have been investigated in detail,
but all these previous investigations rely on the symmetry requirement that
in the background universe both the metric and affine connection are homogeneous and isotropic.
In this paper we relax the symmetry constraint on the connection and leave it arbitrary at the beginning,
after all the cosmological principle only needs the metric of the background spacetime to meet the symmetry requirement.
We find a new flat universe solution for the Nieh-Yan modified teleparallel gravity,
for which the background dynamics itself is unchanged but the perturbations around it
present a new feature that the scalar and tensor perturbations are coupled together at the linear level.
The implications of this peculiar feature in primordial perturbations from inflation are also discussed.
\end{abstract}

\maketitle

\section{introduction}

Stimulated by the experimental detections of gravitational waves (GWs) \cite{ligo1,ligo2}
and the developments in the cosmic microwave background radiation (CMB) experiments \cite{CMB1,CMB2},
parity violating gravities attracted lots of interests in recent years.
A famous and frequently studied parity violating gravity model is the so-called Chern-Simons modified gravity \cite{CSgravity1,CSgravity2}
which within the framework of Riemannian geometry modifies general relativity (GR) by a gravitational Chern-Simons term.
The Chern-Simons modified gravity predicts the difference between the amplitudes of the left- and right-handed polarized components of gravitational waves, i.e., the so-called amplitude birefringence phenomenon.
However, this model was found to suffer from the problem of vacuum
instability because one of the circularly polarized components of GWs becomes a ghost at high frequencies \cite{CSgravity3}.
Further extensions~\cite{Crisostomi:2017ugk,Gao:2019liu,Zhao:2019xmm} to this model did not circumvent this difficulty because in these extended models the pathological behavior still appear at high energy scales,
as shown in Ref. \cite{Bartolo:2020gsh}. It is very difficult to have a ghost-free parity violating gravity model within the framework of Riemannian geometry.

Successful parity violating gravity models are available if we go beyond the Riemannian geometry. For example,
the Nieh-Yan modified teleparallel gravity (NYTG) \cite{PVtele1,PVtele2} is constructed within the framework of the teleparallel gravity (TG) \cite{Tele,tele2021}, where
the gravity is identified with the spacetime torsion in stead of the curvature.
One may have a GR equivalent TG model \cite{Maluf:2013gaa} (we may call it TGR).
The NYTG model \cite{PVtele1,PVtele2} modifies TGR slightly by the anomalous coupling $\theta \mathcal{T}\widetilde{\mathcal{T}}$
between an axion-like field $\theta(x)$ and the Nieh-Yan density \cite{Nieh:1981ww}:
$\mathcal{T}\widetilde{\mathcal{T}}\equiv (1/2)\varepsilon^{\mu\nu\rho\sigma}\mathcal{T}^{\lambda}_{~\mu\nu}\mathcal{T}_{\lambda\rho\sigma}$, where
$\mathcal{T}^{\lambda}_{~\mu\nu}$ is the torsion tensor, $\varepsilon^{\mu\nu\rho\sigma}$ is Levi-Civita tensor which relates the totally antisymmetric symbol
$\epsilon^{\mu\nu\rho\sigma}$ and the determinant of the metric $g$ through the equation $\varepsilon^{\mu\nu\rho\sigma}=\epsilon^{\mu\nu\rho\sigma}/\sqrt{-g}$.
The Nieh-Yan density is parity-odd, so at the background with $\partial_{\mu}\theta\neq 0$,
the Nieh-Yan coupling term $\theta \mathcal{T}\widetilde{\mathcal{T}}$ violates the parity symmetry spontaneously.
The NYTG model has been applied to cosmology in Refs.~\cite{PVtele1,PVtele2}, where
it was found that this model predicts a difference between the propagating velocities of the left- and right-handed polarized components of GWs, i.e.,
the so-called velocity birefringence phenomenon.
More importantly, through detailed investigations on the cosmological perturbations, it was shown in Refs.~\cite{PVtele1,PVtele2} that the NYTG model is ghost-free.
Recently, this model was found to be compatible with the results of most local tests in the Solar System
at the post-Newtonian order \cite{Rao:2021azn,Qiao:2021fwi},
the upper limit on its model parameters by the GWs data of LIGO/Virgo Collaboration was obtained in Ref.~\cite{Wu:2021ndf},
and the enhancement of primordial GWs during inflation due to the velocity birefringence of NYTG model
and its implications in the air-based GWs experiments were studied in Ref.~\cite{Cai:2021uup}.
Other recent studies on parity violating gravities can be found in Refs.~\cite{Li:2021mdp,Gong:2021jgg,Hohmann:2022wrk,Tong:2022cdz,Li:2022vtn,Zhang:2022xmm,Zhu:2022dfq,Zhu:2022uoq,Filho:2022yrk,Qiao:2022mln,Cai:2022lec,Chen:2022wtz}.

In all the previous studies of the cosmological applications of the NYTG model, both the metric and the affine connection of the background universe are required to be homogeneous and isotropic at the beginning.
The spacetime under this strong symmetry constraint is called the regular universe in this paper. The background solutions of the regular universe have been well studied within the TG framework \cite{Hohmann:2019nat,Hohmann:2020zre,Coley:2022qug},
and are universally applicable to almost all TG models.
In fact these solutions have been frequently adopted by different authors, e.g.,
\cite{Myrzakulov:2010vz,Cai:2011tc,Cai:2015emx,Hohmann:2022wrk}
\footnote{
Actually, the cosmological background solution whose tetrad is $e^{A}_{~\mu}=\mathrm{diag}(1,a,a,a)$ or $e^{A}_{~\mu}=\mathrm{diag}(a,a,a,a)$
under the Weitzenb\"{o}ck gauge is the regular flat universe.
However, most of the earlier literature did not clearly point out that
the selection of such a tetrad under the Weitzenb\"{o}ck gauge actually requires the connection to satisfy the same symmetry of the metric.}.

However, the cosmological principle only needs the metric of the background universe to meet the high symmetry requirement. In the Riemannian geometry, once we impose this symmetry requirement on the metric, the connection (i.e., the Christoffel symbol) satisfies the same symmetry requirement automatically.
In TG models, the symmetry constraint on the affine connection is independent of the one on the metric. If one drops this extra constraint on the connection and leaves it arbitrary at the beginning, there will be final solutions for which the connection is neither homogeneous nor isotropic. We call the universe which has a homogeneous and isotropic metric and a non-homogeneous and non-isotropic affine connection the irregular universe.
So far the irregular universe has rarely aroused research interest,
only a few literatures have initially studied the flat irregular universe (or irregular Minkowski spacetime) solutions in $f(\mathbb{T})$ gravity models \cite{Bejarano:2017akj,Bejarano:2019fii,BeltranJimenez:2020fvy,Golovnev:2020nln}.
The irregular universe does not violate the cosmological principle, but questions are in coming:
What features and new physical phenomena could exist in the irregular universe?
Or might the irregular universe have properties that are clearly contradictory to experiments so that only the regular universe is physically feasible?
These questions deserve detailed studies for any TG models.

In this paper, we will study the irregular universe in the NYTG model.
Firstly, we will obtain a more general flat universe solution than those in Refs.~\cite{PVtele1,PVtele2}
by solving the equations of motion of the NYTG model directly
under the condition that only the metric is required to be homogeneous and isotropic.
By analyzing the symmetry of the connection,
we will show that the flat universe we obtain is generally an irregular flat universe, and in special cases it reduces back to a regular universe.
We will also show that even in the irregular flat universe, the background equations in the NYTG model are exactly the same as those in GR.
Secondly, we will study the linear cosmological perturbations around the irregular flat universe.
We will find that tensor perturbations and scalar perturbations are coupled at the linear perturbation level.
This is a peculiar feature that distinguishes the irregular universe from the regular universe in the NYTG model.
We speculate that this peculiar feature is caused by the fact that
the interior space does not satisfy the homogeneity and isotropy in the irregular universe.
Finally, we will study the primordial fluctuations generated by slow-roll inflation in the regular and irregular flat universes.
We will show that the primordial fluctuations of left- and right-handed GWs are different whether in the regular universe or in the irregular universe.
We will also show that there is a strong statistical correlation between primordial scalar fluctuations and primordial tensor fluctuations
generated by slow-roll inflation in the irregular universe.

This paper is organized as follows. In Sec.~\ref{TG and NYTG}, we briefly introduce the TG theory and the NYTG model.
In Sec.~\ref{Nonregular flat universe}, we study spatially flat cosmological background solutions
that only requires the metric to be homogeneous and isotropic in the NYTG model.
In Sec.~\ref{Perturbations}, through the quadratic actions for scalar, vector, and tensor perturbations,
we investigate linear perturbations around the regular and irregular flat universes.
In Sec.~\ref{power spectrum}, we apply our result to the early universe and discuss briefly the primordial  perturbations
generated by slow-roll inflation.

In this paper, we adopt the unit $8\pi G=1$, and use the signature $(+,-,-,-)$ for the metric.
The tensor indices of the interior space are denoted by $A,B,C,...=0, 1, 2, 3$ and by $a, b, c,...=1, 2, 3$ when limiting to spatial components.
They are lowered and raised by the Minkowski metric $\eta_{AB}$ and its inverse $\eta^{AB}$. The spacetime tensor indices are denoted by
Greek $\mu, \nu, \rho,...=0, 1, 2, 3$ and by Latin $i, j, k,...=1, 2, 3$ when limiting to spatial components. They are lowered and raised by the spacetime metric
$g_{\mu\nu}$ and its inverse $g^{\mu\nu}$.
The antisymmetric symbol $\epsilon^{\mu\nu\rho\sigma}$ has the properties: $\epsilon^{0ijk}=\epsilon^{ijk}\equiv\epsilon_{ijk}$, and $\epsilon^{123}=1$.
In addition, we distinguish the spacetime affine connection $\hat{\Gamma}^{\rho}_{~\mu\nu}$
and its associated covariant derivative $\hat{\nabla}$ from the Levi-Civita connection ${\Gamma}^{\rho}_{~\mu\nu}$
and its associated covariant derivative ${\nabla}$ respectively.

\section{TG theory and the NYTG model}\label{TG and NYTG}

The TG theory can be considered as a constrained metric-affine theory.
It is formulated in a spacetime endowed with a metric $g_{\mu\nu}$ and an affine connection $\hat{\Gamma}^{\rho}_{~\mu\nu}$,
which is curvature free and metric compatible,
\be\label{TGconstrain}
\hat{R}^{\sigma}_{~\rho\mu\nu}=\partial_{\mu}\hat{\Gamma}^{\sigma}_{~\nu\rho}-\partial_{\nu}\hat{\Gamma}^{\sigma}_{~\mu\rho}
+\hat{\Gamma}^{\sigma}_{~\mu\lambda}\hat{\Gamma}^{\lambda}_{~\nu\rho}-\hat{\Gamma}^{\sigma}_{~\nu\lambda}\hat{\Gamma}^{\lambda}_{~\mu\rho}=0~,~
\hat{\nabla}_{\rho}g_{\mu\nu}=\partial_{\rho}g_{\mu\nu}
-\hat{\Gamma}^{\lambda}_{~\rho\mu}g_{\lambda\nu}-\hat{\Gamma}^{\lambda}_{~\rho\nu}g_{\mu\lambda}=0~.
\ee
Without curvature and nonmetricity, in the TG theory
the gravity is identified with spacetime torsion $\mathcal{T}^{\rho}_{~\mu\nu}=2\hat{\Gamma}_{\,[\mu\nu]}^{\rho}$.
One can also describe the TG theory using the language of the tetrad $e^{A}_{~\mu}$ and the spin connection $\omega^{A}_{~B\mu}$.
They relates the metric $g_{\mu\nu}$ and the affine connection $\hat{\Gamma}^{\rho}_{~\mu\nu}$ through the following relations
\be\label{metrictetradrelation}
g_{\mu\nu}=\eta_{AB}e^{A}_{~\mu}e^{B}_{~\nu}~~,~~
\hat{\Gamma}^{\rho}_{~\mu\nu}=e_{A}^{~\,\rho}(\partial_{\mu}e^A_{~\nu}+\omega^A_{~B\mu}e^B_{~\nu})~.
\ee
The torsion tensor is written as
\be\label{torsion tensor}
\mathcal{T}^{\rho}_{~\mu\nu}=2e_{A}^{~\,\rho}(\partial_{[\mu}e^{A}_{~\nu]}+\omega^{A}_{~B[\mu}e^{B}_{~\nu]})~.
\ee
The teleparallel constraints (\ref{TGconstrain}) dictate that the spin connection can be in general expressed as
\be\label{omega}
\omega_{~B \mu}^{A}=(\Lambda^{-1})^{A}_{~C} \partial_{\mu} \Lambda_{~B}^{C}~,
\ee
where $\Lambda^{A}_{~B}$ is arbitrary element of Lorentz transformation matrix which is position dependent and satisfies the relation
$\eta_{AB}\Lambda^A_{~C}\Lambda^B_{~D}=\eta_{CD}$ at any spacetime point.
Therefore, the tetrad $e^{A}_{~\mu}$ and the Lorentz matrix $\Lambda^{A}_{~B}$ can be regarded as the basic variables of the TG theory.
In this way, the teleparallel constraints (\ref{TGconstrain}) are automatically satisfied.

The TGR model, as the GR equivalent TG model, has the following action,
\bea\label{TEGRaction}
S_{TGR}=\frac{1}{2}\int \zd^4x ~{|e|}\,\mathbb{T}\equiv\int \zd^4x~{|e|}
\left(-\frac{1}{2}\mathcal{T}_{\mu}\mathcal{T}^{\mu}+\frac{1}{8}\mathcal{T}_{\alpha\beta\mu}\mathcal{T}^{\alpha\beta\mu}
+\frac{1}{4}\mathcal{T}_{\alpha\beta\mu}\mathcal{T}^{\beta\alpha\mu}\right)~,
\eea
where ${ |e|}=\sqrt{-g}$ is the determinant of the tetrad, $\mathbb{T}$ is the torsion scalar, and
$\mathcal{T}_{\mu}=\mathcal{T}^{\alpha}_{~~\mu\alpha}$ is the
torsion vector.
Since we have the identity $-{R}(e)=\mathbb{T}+2{\nabla}_{\mu}\mathcal{T}^{\mu}$,
the action (\ref{TEGRaction}) is identical to the Einstein-Hilbert action up to a surface term,
where the curvature scalar $R(e)$ is defined by the Levi-Civita connection and considered
as being fully constructed from the metric, and in turn from the tetrad.
Since the surface term in the action does not affect the equations of motion,
we say that the TGR is equivalent to GR at the level of the equations of motion.

The NYTG model \cite{PVtele1,PVtele2} modifies the TGR model by introducing the coupling
\be\label{NY}
S_{NY}=\frac{c}{4}\int \zd^4x ~{|e|} ~\theta\,\mathcal{T}\widetilde{\mathcal{T}}~,
\ee
between an axion-like field $\theta$ and the Nieh-Yan density $\mathcal{T}\widetilde{\mathcal{T}}$. The coupling constant $c$ is dimensionless.
Generally we should also consider its own dynamics of the axion-like field and take other matter into account, so the full action of the NYTG model is
\bea\label{NYTG}
S_{NYTG}=\int \zd^4x~ {|e|}\left[\frac{1}{2}\mathbb{T}
+\frac{c}{4}\,\theta\,\mathcal{T}\widetilde{\mathcal{T}}+\frac{1}{2}\nabla_{\mu}\theta\nabla^{\mu}\theta-V(\theta)\right]+S_m~.
\eea
Other matter with the action $S_m$ is assumed to be coupled to spacetime minimally through the tetrad. At the background in which the axion-like field has
non-zero spacetime derivatives, the Nieh-Yan coupling term breaks parity spontaneously.
Because only the first-order derivatives of the basic variables appears in the action,
the NYTG model can avoid the Ostrogradski ghost mode, which is expected to be originated from higher-order derivatives in the action \cite{Woodard:2006nt}.

As with most modified TG theories,
the NYTG model apparently has two kinds of gauge symmetries: diffeomorphism invariance and local Lorentz invariance.
The latter transformation makes the following change:
\be\label{LT}
e^{A}_{~\mu}\rightarrow(L^{-1})^{A}_{~B}e^{B}_{~\mu}~,~ \Lambda^{A}_{~B}\rightarrow\Lambda^{A}_{~C}L^{C}_{~B}~,
\ee
where $L^{A}_{~B}(x)$ are the element of Lorentz matrix.
We would like to use different notations to distinguish two kinds of Lorentz matrices:
$\Lambda^{A}_{~B}(x)$ is used to express the spin connection as in Eq.~(\ref{omega}),
but $L^{A}_{~B}(x)$ represents the local transformation that makes a shift from one local frame to another.
Transformation (\ref{LT}) can be expressed in terms of tetrad and spin connections as
\be\label{LT2}
e^{A}_{~\mu}\rightarrow(L^{-1})^{A}_{~B}e^{B}_{~\mu}~,~
\omega^{A}_{~B\mu}\rightarrow (L^{-1})^{A}_{~C}\omega^{C}_{~D\mu}L^{D}_{~B}+(L^{-1})^{A}_{~C}\partial_{\mu}L^{C}_{~B}~.
\ee
It is easy to prove that the metric $g_{\mu\nu}$ and torsion tensor $\mathcal{T}_{~\mu \nu}^{\rho}$
are invariant under the local Lorentz transformation (\ref{LT}), as is the action (\ref{NYTG}).
Due to the local Lorentz invariance, one can choose the gauge $\Lambda^{A}_{~B}=\delta^{A}_{~B}$, i.e., $\omega^{A}_{~B\mu}=0$.
This is the Weitzenb\"{o}ck connection, which has been frequently adopted in the literature.
In addition, there is another symmetry hidden in the NYTG model.
The Nieh-Yan term (\ref{NY}) can be integrated by parts as
\be\label{NY2}
S_{NY}=-\frac{c}{2}\int \zd^4x~\eta_{AB}\epsilon^{\mu\nu\rho\sigma}(\partial_{\mu}\theta)
(\Lambda^{A}_{~C}e^{C}_{~\nu})\partial_{\rho}(\Lambda^{B}_{~D}e^{D}_{~\sigma})~.
\ee
It can be seen that the Nieh-Yan term (\ref{NY}) is invariant under the following transformation
\be\label{trans0}
(\Lambda^{A}_{~C}e^{C}_{~\mu})\rightarrow L^{A}_{~B}(\theta)(\Lambda^{B}_{~C}e^{C}_{~\mu})\, ,
\ee
where $L^{A}_{~B}(\theta)$ is Lorentz matrix that depends only on axion-like field $\theta$.
Note that $\Lambda^{A}_{~C}e^{C}_{~\mu}$ is invariant under transformation (\ref{LT}).
Due to the Lorentz symmetry (\ref{LT}), the transformation (\ref{trans0}) can always be attributed to the fact that
the tetrad $e^{A}_{~\mu}$ remains unchanged while the Lorentz matrix $\Lambda^{A}_{~B}$ undergoes a Lorentz transformation.
Obviously the metric and the action of TGR model are invariant under such a transformation.
So the total action of the NYTG model is invariant under the transformation (\ref{trans0}).

The equations of motion follow from the variation of the action (\ref{NYTG}) with respect to $e^{A}_{~\mu}$ and $\Lambda^{A}_{~B}$ separately
\bea
 G^{\mu\nu}+N^{\mu\nu}&=&T^{\mu\nu}+T^{\mu\nu}_{\theta}~,\label{eom1}\\
 N^{[\mu\nu]}&=&0~,\label{eom2}
\eea
where $N^{\mu \nu}=(c/2)\varepsilon^{\mu\lambda\rho\sigma}\partial_{\lambda}\theta\,\mathcal{T}^{\nu}_{~\rho\sigma}$,
$G^{\mu\nu}$ is the Einstein tensor, $T^{\mu\nu}=-(2/\sqrt{-g})(\delta S_m/\delta g_{\mu\nu})$
and $T^{\mu\nu}_{\theta}=[V(\theta)-\nabla_{\alpha}\theta\nabla^{\alpha}\theta/2]g^{\mu\nu}+\nabla^{\mu}\theta\nabla^{\nu}\theta$
are the energy-momentum tensors for the matter and the axion-like field $\theta$ respectively.
Similar to most modified TG models, the equation of motion (\ref{eom2}) from the variation of $\Lambda^{A}_{~B}$ is
not independent of Eq.~(\ref{eom1}), it is just the antisymmetric part of the latter.
As explained in Ref.~\cite{PVtele2}, this is due to the local Lorentz invariance of the action,
any change caused by $\delta\Lambda^{A}_{~B}$ can always be equivalent to the change caused by $\delta e^{A}_{~\mu}$,
so requiring the action to take the extremum under $\delta e^{A}_{~\mu}$
already includes the case where the action takes the extremum under $\delta\Lambda^{A}_{~B}$.
There is another equation following from the variation of the action (\ref{NYTG}) with respect to $\theta$,
\be\label{eom3}
{\square}\theta+V^{(1)}-\frac{c}{4} \mathcal{T}\widetilde{\mathcal{T}}=0~,
\ee
where ${\square}=g^{\mu\nu}{\nabla}_{\mu}{\nabla}_{\nu}$ and $V^{(n)}=\zd^{n}V(\theta)/\zd\theta^{n}$.
All of these equations of motion are consistent with the Bianchi identity $\nabla_{\mu}G^{\mu\nu}=0$ and
the covariant conservation law $\nabla_{\mu} T^{\mu \nu}=0$.

Also in Refs.~\cite{PVtele1,PVtele2}, the cosmological perturbations of the NYTG model were analyzed in detail.
It was found that the NYTG model makes a difference between the
propagating velocities of the left- and right-handed polarized components of GWs, but makes no difference between their amplitudes.
This phenomenon is called velocity birefringence, which is a clear physical signal of parity violation.
More importantly, the NYTG model was confirmed to be ghost free through the quadratic action of cosmological perturbations.

It is worth mentioning that the Nieh-Yan density $\mathcal{T}\widetilde{\mathcal{T}}$ is not the only parity-odd term within the TG framework. A more general
model including all the parity-odd terms which are quadratic in the torsion tensor was considered in Ref.~\cite{PVtele3}.
But then it was found in Ref.~\cite{PVtele4} that this more general model suffers from the problem of ghost instability again,
unless it completely reduces to the NYTG model.
Therefore, within the TG framework, for all parity-odd terms which are quadratic in the torsion tensor,
only the Nieh-Yan density $\mathcal{T}\widetilde{\mathcal{T}}$ can avoid the ghost instability.
This means the NYTG model is robust to some extent.

\section{Irregular flat universe in the NYTG model}\label{Nonregular flat universe}

So far all the studies on the cosmological applications of the NYTG model only considered the regular universe as the background, that means both the metric and the affine connection are constrained to be homogeneous and isotropic. This constraint may be too strong, after all the cosmological principle which is supported by current observations only needs the metric of the background spacetime to meet the high symmetry requirement.
In this paper,  we will drop the symmetry requirement on the connection and leave it arbitrary at the beginning. After this relaxation, it is expected that
the NYTG model will have more interesting cosmological background solutions. We are interested in the irregular universe solutions in which the metric homogeneous and isotropic but the connection is neither homogeneous nor isotropic. For simplicity, we will only consider the spatially flat universe.

In flat universe, the metric can be expressed in rectangular coordinate as
\be\label{FRWmetric}
\zd s^{2}=g_{\mu\nu}\zd x^{\mu}\zd x^{\nu}=a^{2}\left(\zd\eta^{2}-\delta_{ij} \zd x^{i} \zd x^{j}\right)~,
\ee
where $a=a(\eta)$ is the scale factor of the universe, $\eta$ is the conformal time.
This is the Friedmann-Robertson-Walker (FRW) metric.
There are 6 Killing vector fields $\{\xi_{I}^{\mu}, I=1,2...6\}$ in flat universe, which can be expressed as
\be\label{killingvector}
\xi^{\mu}_{I}=\delta^{\, \mu}_{~I}~,~
\xi^{\mu}_{I+3}=\epsilon_{Iij}\delta^{\mu}_{~i}x^{j}~,~~~~ I=1,2,3
\ee
where $\xi^{\mu}_{1}, \xi^{\mu}_{2}, \xi^{\mu}_{3}$ are Killing vector fields representing the symmetry of spatial translation,
and $\xi^{\mu}_{4}, \xi^{\mu}_{5}, \xi^{\mu}_{6}$ are Killing vector fields representing the symmetry of spatial rotation.
One can prove that the FRW metric satisfies the condition: $\mathcal{L}_{\xi_{I}}g_{\mu\nu}=0$,
where $\mathcal{L}_{\xi_{I}}$ is the Lie derivative along the Killing vector field $\xi^{\mu}_{I}$.
This reflects the result that the metric is homogeneous and isotropic.
One can also prove that $\mathcal{L}_{\xi_{I}}\Gamma^{\rho}_{~\mu\nu}=0$ for the Levi-Civita connection $\Gamma^{\rho}_{~\mu\nu}$, which is automatically homogeneous and isotropic.
This is why we do not need to pay extra attention to the symmetry of the connection within the framework of Riemannian geometry.

\subsection{Regular flat universe}

For TG models, even the metric is determined, the affine connection is still arbitrary to some extent.
Usually, as suggested in Refs~\cite{Hohmann:2019nat,Hohmann:2020zre,Coley:2022qug}, a further constraint was imposed that requires the connection is also homogeneous and isotropic, that is,
\be\label{FRWconnection}
\mathcal{L}_{\xi_I}\hat{\Gamma}^{\rho}_{~\mu\nu}=\hat{\nabla}_{\mu}\hat{\nabla}_{\nu}\,\xi^{\rho}_{I}
-\hat{\nabla}_{\mu}(\mathcal{T}^{\rho}_{~\nu\sigma}\xi^{\sigma}_{I})=0~.
\ee
Although $\hat{\Gamma}^{\rho}_{~\mu\nu}$ is coordinate dependent, the Lie derivative of $\hat{\Gamma}^{\rho}_{~\mu\nu}$ does not depend on the coordinate.
Hence the condition (\ref{FRWconnection}) is unambiguous.
Combining Eqs.~(\ref{FRWmetric}) and (\ref{FRWconnection}) selected the regular flat universe solution in which the tetrad $e^{A}_{\mu}$ and Lorentz matrix $\Lambda^{A}_{~B}$ have the following forms:
\be\label{regular}
e^{A}_{~\mu}=a\delta^{A}_{~\mu}~,~\Lambda^{A}_{~B}=\mathring{\Lambda}^{A}_{~B}~,
\ee
where $\mathring{\Lambda}^{A}_{~B}$ is a global Lorentz matrix, which does not depend on spacetime.
All other solutions satisfying Eqs.~(\ref{FRWmetric}) and (\ref{FRWconnection})
differ from the solution (\ref{regular}) only by Lorentz transformation (\ref{LT}), so they are physically equivalent to the solution (\ref{regular}).
The above process does not depend on a specific TG theory, so the solution (\ref{regular}) is generally applicable to most TG theories.

For the NYTG model, the solution (\ref{regular}) can automatically satisfy the constraint $N^{[\mu\nu]}=0$,
so the solution (\ref{regular}) is compatible with the NYTG model.
Furthermore, solution (\ref{regular}) leads to $N^{\mu\nu}=0$ and $\mathcal{T}\widetilde{\mathcal{T}}=0$,
which means that the Nieh-Yan term has no effect on the regular flat universe background.
Therefore, the background equations of the regular flat universe are exactly the same as those of GR \cite{PVtele1,PVtele2}.

\subsection{Irregular flat universe}

To look for the irregular universe solution, we should give up the constraint (\ref{FRWconnection}) on the connection. After this relaxation, the connection is left to be determined by the equation of motion.

In a flat universe, we can always simply find the non-zero components of $G^{\mu\nu}$, $T^{\mu\nu}$ and $T^{\mu\nu}_{\theta}$ as
\be\label{nonvanish}
G^{00}=\frac{3\mathcal{H}^{2}}{a^{4}}~,~T^{00}=\frac{\rho}{a^{2}}~,~T_{\theta}^{00}=\frac{\rho_{\theta}}{a^{2}}~,~
G^{ij}=-\frac{2\mathcal{H}'+\mathcal{H}^{2}}{a^{4}}\delta_{ij}~,~T^{ij}=\frac{p}{a^{2}}\delta_{ij}~,~T_{\theta}^{ij}=\frac{p_{\theta}}{a^{2}}\delta_{ij}~,
\ee
where $\mathcal{H}=a'/a$ is the conformal Hubble rate, prime represents the derivative with respect to the conformal time $\eta$,
$\rho_{\theta}=\theta^{\prime 2} /\left(2 a^{2}\right)+V$ and $p_{\theta}=\theta^{\prime 2} /\left(2 a^{2}\right)-V$
are the energy density and pressure of the $\theta$ field, and $\rho$ and $p$ denote the energy density and pressure of other matter.
Thanks to the Lorentz symmetry (\ref{LT}), we can always reduce the tetrad to the simple form $e^{A}_{~\mu}=a\delta^{A}_{~\mu}$ in flat universe.
In order to facilitate further analysis, we decompose the independent non-zero components of spin connections $\omega^{A}_{~B\mu}$ as follows
\bea\label{spindecompose}
& &\nonumber
\delta^{a}_{~i}\omega^{0}_{~a0}=\mU_{i}~~,~~
\delta^{i}_{~a}\delta^{b}_{~j}\omega^{a}_{~bk}=\Sigma\epsilon_{ijk}+\epsilon_{ijl}\Sigma_{kl}+\Sigma_{i}\delta_{jk}-\Sigma_{j}\delta_{ik}~,
\\
& &
\delta^{i}_{~a}\delta^{b}_{~j}\omega^{a}_{~b0}=\epsilon_{ijk}\mV_{k}~~,~~
\delta^{a}_{~i}\omega^{0}_{~aj}=\sigma\delta_{ij}+\sigma_{ij}+\epsilon_{ijk}\sigma_{k}~,
\eea
where $\Sigma_{ij}$ and $\sigma_{ij}$ are symmetric and traceless spatial tensors.
In the above decomposition we have exploited the property $\omega_{AB\mu}=-\omega_{BA\mu}$ due to $\hat{\nabla}_{\rho}g_{\mu\nu}=0$.
Note that the variables $\sigma, \Sigma, \mU_{i}, \mV_{i}, \sigma_{i}, \Sigma_{i}, \sigma_{ij}, \Sigma_{ij}$
are not completely independent because we have not yet imposed $\hat{R}^{\sigma}_{~\rho\mu\nu}=0$ on the spin connection.
Combining $e^{A}_{~\mu}=a\delta^{A}_{~\mu}$  and Eq.~(\ref{spindecompose}), $N^{\mu\nu}$ can be obtained as
\be
N^{00}=0~~,~~N^{0i}=0~~,~~N^{i0}=\frac{2c\theta'}{a^{4}}\sigma_{i}~~,
~~N^{ij}=\frac{c\theta'}{a^{4}}\left(2\Sigma\delta_{ij}-\Sigma_{ij}+\epsilon_{ijk}\Sigma_{k}\right)~.
\ee
In order for Eqs.~(\ref{eom1}) and (\ref{eom2}) to hold, there must be
\be\label{condition1}
\sigma_{i}=0~~,~~\Sigma_{i}=0~~,~~\Sigma_{ij}=0~~,~~\Sigma=\Sigma(\eta)~.
\ee
Combining $e^{A}_{~\mu}=a\delta^{A}_{~\mu}$, Eqs.~(\ref{spindecompose}) and (\ref{condition1}), Nieh-Yan density can be obtained as
\be\label{NYdensity2}
\mathcal{T}\widetilde{\mathcal{T}}=\frac{24\Sigma}{a^{2}}(\mathcal{H}-\sigma)~.
\ee
In order for Eq.~(\ref{eom3}) to hold, the Nieh-Yan density $\mathcal{T}\widetilde{\mathcal{T}}$ can only be a function of time $\eta$, so
$\sigma=\sigma(\eta)$ when $\Sigma\neq0$.

Combining Eqs.~(\ref{spindecompose}) and (\ref{condition1}),
$\hat{R}^{\sigma}_{~\rho\mu\nu}=0$ gives
\bea
& & \label{cureom1}
\mathcal{S}'_{ij}-\mU_{i,j}+\epsilon_{ijk}\Sigma\, \mU_{k}+ \epsilon_{ikl}\mathcal{S}_{jk}\mV_{l}=0~, \\
& & \label{cureom2}
\Sigma'\delta_{ij}-\mV_{i,j}+\epsilon_{ijk}\Sigma\, \mU_{k}- \epsilon_{ikl}\mathcal{S}_{jk}\mU_{l}=0~, \\
& & \label{cureom3}
\epsilon_{ikl}\mathcal{S}_{lj,k}+\Sigma(\mathcal{S}_{ij}-\mathcal{S}_{kk}\delta_{ij})=0~,\\
& & \label{cureom4}
\epsilon_{inm}\mathcal{S}_{jn}\mathcal{S}_{km}-\Sigma^{2}\epsilon_{ijk}=0~,
\eea
where $\mathcal{S}_{ij}=\sigma\delta_{ij}+\sigma_{ij}$ and the subscript ``$,i$" represents a derivative with respect to $x^{i}$.
The trace of Eq.~(\ref{cureom3}) gives
\be\label{cureom3o1}
\sigma \Sigma=0~.
\ee
This means that at least one of $\sigma$ and $\Sigma$  is zero.
If $\sigma=0$,
the equation after the Hodge duality of the "$j,k$" index in Eq.~(\ref{cureom4})
can be decomposed as follows according to the trace part and the traceless part:
\be\label{cureom4o1}
6\,\Sigma^{2}+\sigma_{ij}\sigma_{ij}=0~~,~~\sigma_{ik}\sigma_{jk}-\frac{1}{3}(\sigma_{kl}\sigma_{kl})\delta_{ij}=0~.
\ee
The solution of Eq.~(\ref{cureom4o1}) is $\Sigma=0, \sigma_{ij}=0$.
This means that  Eqs.~(\ref{cureom4}) and (\ref{cureom3o1}) must give
\be\label{cureom5}
\Sigma=0~.
\ee

Combining Eqs.~(\ref{condition1}) and (\ref{cureom5}) gives $N^{\mu\nu}=0$ and $\mathcal{T}\widetilde{\mathcal{T}}=0$,
which means that the Nieh-Yan term has no effect even on the irregular flat universe background.
Therefore, the background equations of the irregular flat universe are exactly the same as those of GR.
This is a somewhat unexpected result.
But the fact that Nieh-Yan term has no effect on the background does not mean that it has no effect on the perturbations.
In order to analyze the perturbations, we need to first find the background solution of the irregular flat universe.

Substituting Eq.~(\ref{cureom5}) into Eqs.~(\ref{cureom1}), (\ref{cureom2}), (\ref{cureom3}) and (\ref{cureom4}), we get
\bea
& & \label{cureom11}
\mathcal{S}'_{ij}-\mU_{i,j}+\epsilon_{ikl}\mathcal{S}_{jk}\mV_{l}=0~, \\
& & \label{cureom22}
\mV_{i,j}+ \epsilon_{ikl}\mathcal{S}_{jk}\mU_{l}=0~, \\
& & \label{cureom33}
\epsilon_{ikl}\mathcal{S}_{lj,k}=0~,\\
& & \label{cureom44}
\epsilon_{inm}\mathcal{S}_{jn}\mathcal{S}_{km}=0~,
\eea
Although there are more equations than variables, this does not mean that
Eqs.~(\ref{cureom11}), (\ref{cureom22}), (\ref{cureom33}) and (\ref{cureom44}) have no solution.
It can be verified that the following are the solution of Eqs.~(\ref{cureom11}), (\ref{cureom22}), (\ref{cureom33}) and (\ref{cureom44})
\bea\label{solution1}
& &\nonumber
\mathcal{S}_{ij}=v_{i}v_{j}f(\eta)F^{(1)}(\vec{v}\cdot\vec{x})~,\\
& &\nonumber
\mV_{i}=g_{a}(\eta)\alpha^{a}_{i}(\eta, \vec{x})-h_{a}(\eta)\beta^{a}_{i}(\eta, \vec{x})~,\\
& &
\mU_{i}=h_{a}(\eta)\alpha^{a}_{i}(\eta, \vec{x})+g_{a}(\eta)\beta^{a}_{i}(\eta, \vec{x})+v_{i} f^{(1)}(\eta)F(\vec{v}\cdot\vec{x})~,
\eea
where
\bea
& &\nonumber
\alpha^{a}_{i}(\eta, \vec{x})=\cosh\left[vf(\eta)F(\vec{v}\cdot\vec{x})\right]\delta_{ai}
+\frac{v_{a}v_{i}}{v^{2}}\Big(1-\cosh\left[vf(\eta)F(\vec{v}\cdot\vec{x})\right]\Big)~,\\
& &\nonumber
\beta^{a}_{i}(\eta, \vec{x})=\epsilon_{aij}\frac{v_{j}}{v}\sinh\left[vf(\eta)F(\vec{v}\cdot\vec{x})\right]~,
\eea
where $v_{1}, v_{2}, v_{3}$ are constant parameters, $v=\sqrt{\delta^{ij}v_{i}v_{j}}$, $\vec{v}\cdot\vec{x}=v_{i}x^{i}$,
$f(\eta), g_{a}(\eta), h_{a}(\eta)$ are arbitrary smooth function of conformal time $\eta$,
$F(\vec{v}\cdot\vec{x})$ is arbitrary smooth function of $\vec{v}\cdot\vec{x}$,
$f^{(n)}(\eta)$ is the $n$ derivative of $f(\eta)$ with respect to conformal time $\eta$,
and $F^{(n)}(\vec{v}\cdot\vec{x})$ is the $n$ derivative of $F(\vec{v}\cdot\vec{x})$ with respect to $\vec{v}\cdot\vec{x}$.

Putting solutions (\ref{condition1}), (\ref{cureom5}) and (\ref{solution1}) into the  decomposition (\ref{spindecompose}),
the spin connection $\omega^{A}_{~B\mu}$ when the tetrad is $e^{A}_{~\mu}=a\delta^{A}_{~\mu}$ can be obtained as
\bea\label{spinconnection1}
& &\nonumber
\omega^{a}_{~00}=\omega^{0}_{~a0}=h_{c}(\eta)\alpha^{c}_{a}(\eta, \vec{x})+g_{c}(\eta)\beta^{c}_{a}(\eta, \vec{x})+v_{a} f^{(1)}(\eta)F(\vec{v}\cdot\vec{x})~,\\
& &\nonumber
\omega^{a}_{~b0}=\epsilon_{abi}\left[g_{c}(\eta)\alpha^{c}_{i}(\eta, \vec{x})-h_{c}(\eta)\beta^{c}_{i}(\eta, \vec{x})\right]~,\\
& &
\omega^{0}_{~ai}=\omega^{a}_{~0i}=v_{a}v_{i}f(\eta)F^{(1)}(\vec{v}\cdot\vec{x})~,~~\omega^{a}_{~bi}=0~.
\eea
It can be verified that the spin connection (\ref{spinconnection1}) does satisfy the teleparallel constraints (\ref{TGconstrain}).
Due to the symmetry (\ref{trans0}), not every $h_{I}(\eta)$ and $g_{I}(\eta)$ represent a physically inequivalent solution.
In order to see this better, we perform a Lorentz transformation (\ref{LT2}) on the above solution.
The transformation matrix $L^{A}_{~B}$ is
\bea\label{LLL}
& &\nonumber
L^{0}_{~0}=\cosh\left[vf(\eta)F(\vec{v}\cdot\vec{x})\right]~,~
L^{0}_{~a}=L^{a}_{~0}=\frac{v_{a}}{v}\sinh\left[vf(\eta)F(\vec{v}\cdot\vec{x})\right]~,\\
& &
L^{a}_{~b}=\delta_{ab}+\frac{v_{a}v_{b}}{v^{2}}\Big(\cosh\left[vf(\eta)F(\vec{v}\cdot\vec{x})\right]-1\Big)~,
\eea
Then, the tetrad $\tilde{e}^{A}_{~\mu}=L^{A}_{~B}e^{B}_{~\mu}$  and the corresponding spin connection $\tilde{\omega}^{A}_{~B\mu}$ are
\bea\label{solution2}
& &\nonumber
\tilde{e}^{0}_{~0}=a\cosh\left[vf(\eta)F(\vec{v}\cdot\vec{x})\right]~,
\tilde{e}^{a}_{~0}=\delta^{ai}\tilde{e}^{0}_{~i}=a\frac{v_{a}}{v}\sinh\left[vf(\eta)F(\vec{v}\cdot\vec{x})\right]~,\\
& &\nonumber
\tilde{e}^{a}_{~i}=a\left[\delta_{ai}+\frac{v_{a}v_{i}}{v^{2}}\big(\cosh\left[vf(\eta)F(\vec{v}\cdot\vec{x})\right]-1\big)\right]~,\\
& &
\tilde{\omega}^{a}_{~00}=\tilde{\omega}^{0}_{~a0}=h_{a}(\eta)~,~
\tilde{\omega}^{a}_{~b0}=\epsilon_{abc}g_{b}(\eta)~,~\tilde{\omega}^{A}_{~Bi}=0~.
\eea
It can be verified that the metric $g_{\mu\nu}$ and connection $\hat{\Gamma}^{\rho}_{~\mu\nu}$
given by solution (\ref{solution2}) are the same as those given by the tetrad $e^{A}_{~\mu}=a\delta^{A}_{~\mu}$ and the spin connection (\ref{spinconnection1}).
Since the solution (\ref{solution2}) satisfies the teleparallel constraints (\ref{TGconstrain}),
the spin connection $\tilde{\omega}^{A}_{~B\mu}$ in the solution (\ref{solution2})
can be expressed by a Lorentz matrix $\tilde{\Lambda}^{A}_{~B}(\eta, \vec{x})$.
And $\tilde{\omega}^{A}_{~Bi}=0$ means that $\tilde{\Lambda}^{A}_{~B}(\eta, \vec{x})=\tilde{\Lambda}^{A}_{~B}(\eta)$.
So taking different $h_{a}(\eta)$ and $g_{a}(\eta)$ is actually taking different $\tilde{\Lambda}^{A}_{~B}(\eta)$.
Since $\theta=\theta(\eta)$ in the cosmological background,
different $\tilde{\Lambda}^{A}_{~B}(\eta)$ can be converted to each other through the Lorentz transformation
$\tilde{\Lambda}^{A}_{~B}(\eta)\rightarrow L^{A}_{~C}(\theta) \tilde{\Lambda}^{C}_{~B}(\eta)$.
Therefore, the solutions with different $h_{a}(\eta)$ and $g_{a}(\eta)$  can be transformed into each other by transformation (\ref{trans0}),
so they are physically equivalent.
In this case, we only need to consider the simplest case below,
that is, the case where $h_{a}(\eta)=g_{a}(\eta)=0$, so that the solution (\ref{spinconnection1}) can be simplified to
\bea\label{solution3}
& &\nonumber
e^{A}_{~\mu}=a\delta^{A}_{~\mu}~,\\
& &\nonumber
\omega^{a}_{~00}=\omega^{0}_{a0}=v_{a} f^{(1)}(\eta)F(\vec{v}\cdot\vec{x})~,~\omega^{a}_{~b0}=0~,\\
& &
\omega^{a}_{~0i}=\omega^{0}_{ai}=v_{a}v_{i} f(\eta)F^{(1)}(\vec{v}\cdot\vec{x})~,~\omega^{a}_{~bi}=0~.
\eea
The solution (\ref{solution3}) can be expressed by the tetrad $e^{A}_{~\mu}$ and the Lorentz matrix $\Lambda^{A}_{~B}$ as
\be\label{solution0}
e^{A}_{~\mu}=a\delta^{A}_{~\mu}~,~~
\Lambda=\mathring{\Lambda}\cdot\exp\Big[f(\eta)F(\vec{v}\cdot\vec{x})\,v_{a}\mathbb{K}^{a}\Big]~,
\ee
where $\mathring{\Lambda}$ is a spacetime independent Lorentz matrix,
and $\mathbb{K}^{1}, \mathbb{K}^{2}, \mathbb{K}^{3}$ are the boost matrices whose expression are
\be\nonumber
\mathbb{K}^{1}=
\left(\begin{array}{cccc}0&1&0&0\\1&0&0&0\\0&0&0&0\\0&0&0&0\end{array}\right)~,~~
\mathbb{K}^{2}=
\left(\begin{array}{cccc}0&0&1&0\\0&0&0&0\\1&0&0&0\\0&0&0&0\end{array}\right)~,~~
\mathbb{K}^{3}=
\left(\begin{array}{cccc}0&0&0&1\\0&0&0&0\\0&0&0&0\\1&0&0&0\end{array}\right)~.
\ee
Regardless of the functional form of $f(\eta)$ and $F(\vec{v}\cdot\vec{x})$,
it can be verified that the solution (\ref{solution0}) always satisfies the teleparallel constraints (\ref{TGconstrain})
and makes Eqs.~(\ref{eom1}) and (\ref{eom3}) self-consistent.
Note that the boosted tetrad solutions similar to the solution (\ref{solution0})
are also discussed in Ref.~\cite{Bejarano:2019fii,BeltranJimenez:2020fvy,Golovnev:2020nln}.
Putting solution (\ref{solution0}) into Eqs.~(\ref{eom1}) and (\ref{eom3}),  we can get
\be\label{backgroundeom}
3 \mathcal{H}^{2}=a^{2}\left(\rho_{\theta}+\rho\right)~,~~
2 \mathcal{H}^{\prime}+\mathcal{H}^{2}=-a^{2}\left(p_{\theta}+p\right)~,~~
\theta^{\prime \prime}+2 \mathcal{H} \theta^{\prime}+a^{2} V^{(1)}=0~.
\ee
The background equations are exactly the same as those of GR.
This means that the Nieh-Yan term has no effect even on the irregular flat universe background.
This is consistent with our analysis above.

Finally, let's focus on the symmetry of the connection given by the solution (\ref{solution0}).
The non-zero components of $\mathcal{L}_{\xi_I}\hat{\Gamma}^{\rho}_{~\mu\nu}$ given by the solution (\ref{solution0}) are
\bea\label{LieG}
& &\nonumber
\mathcal{L}_{\xi_I}\hat{\Gamma}^{0}_{~0i}=\mathcal{L}_{\xi_I}\hat{\Gamma}^{i}_{~00}=v_{I}v_{i}f^{(1)}(\eta)F^{(1)}(\vec{v}\cdot\vec{x})~,\\
& &\nonumber
\mathcal{L}_{\xi_I}\hat{\Gamma}^{0}_{~ij}=\mathcal{L}_{\xi_I}\hat{\Gamma}^{i}_{~j0}=v_{I}v_{i}v_{j}f(\eta)F^{(2)}(\vec{v}\cdot\vec{x})~,\\
& &\nonumber
\mathcal{L}_{\xi_{I+3}}\hat{\Gamma}^{0}_{~0i}=\mathcal{L}_{\xi_{I+3}}\hat{\Gamma}^{i}_{~00}=
-\epsilon_{Iij}v_{j}f^{(1)}(\eta)F(\vec{v}\cdot\vec{x})+v_{i}\epsilon_{Ijk}v_{j}x^{k}f^{(1)}(\eta)F^{(1)}(\vec{v}\cdot\vec{x})~,\\
& &
\mathcal{L}_{\xi_{I+3}}\hat{\Gamma}^{0}_{~ij}=\mathcal{L}_{\xi_{I+3}}\hat{\Gamma}^{i}_{~j0}=
2v_{(i}\epsilon_{j)Ik}v_{k}f(\eta)F^{(1)}(\vec{v}\cdot\vec{x})+v_{i}v_{j}\epsilon_{Ikl}v_{k}x^{l}f(\eta)F^{(2)}(\vec{v}\cdot\vec{x})~,
\eea
where $I=1, 2, 3$ in Eq.~(\ref{LieG}), and the subscript parentheses denotes the symmetrization.
The fact that $\mathcal{L}_{\xi_I}\hat{\Gamma}^{\rho}_{~\mu\nu}\neq0$
indicates that the spacetime connection given by the solution (\ref{solution0}) is neither homogeneous nor isotropic.
So the solution (\ref{solution0}) does represent an irregular flat universe.
When $v_{i}=0$ or $f(\eta)=0$ or $F(\vec{v}\cdot\vec{x})=0$, there is $\mathcal{L}_{\xi_I}\hat{\Gamma}^{\rho}_{~\mu\nu}=0$,
and the solution (\ref{solution0}) dose reduce to the regular flat universe solution (\ref{regular}).

\section{Perturbations around the irregular flat universe}\label{Perturbations}

In the previous section we studied the flat universe solution of the NYTG model
that only requires the metric to be homogeneous and isotropic.
We found that the Nieh-Yan term has no effect even on the irregular flat universe background.
In order to explore the effect of the Nieh-Yan term on the irregular flat universe,
we study the linear cosmological perturbations around the irregular flat universe (\ref{solution0}) in this section.
For simplicity, we only consider the case of $F(\vec{v}\cdot\vec{x})=\vec{v}\cdot\vec{x}$,
which is equivalent to requiring that the coefficients of the equations of linear perturbations do not depend on the spatial coordinates $\vec{x}$
(see below for details).
And we also ignore other matter so that $S_m=0$.

We use the following parametrization for perturbed tetrad \cite{Izumi:2012qj}:
\bea
& & e^{0}_{\ 0}=a(1+A)~,~\nonumber e^{0}_{\ i}=a(\beta_{,i}+\beta_{i}^{V})~,
~\nonumber e^{c}_{\ 0}=a\delta_{ci}(\chi_{,i}+\chi_{i}^{V})~,\nonumber\\
& & e^{c}_{\ i}=a\delta_{cj}[ (1-\psi)\delta_{ij}+\alpha_{,ij}+\alpha_{j,i}^{V}-
              \epsilon_{ijk}(\lambda_{,k}+\lambda_{k}^{V})+\frac{1}{2}h^{T}_{ij}]~,
\eea
So the perturbed metric components have the familiar forms:
\bea
& &g_{00}=a^{2}(1+2A)~,~ g_{0i}=-a^{2}(B_{,i}+B_{i}^{V})~,\nonumber\\
& &g_{ij}=-a^{2}[(1-2\psi)\delta_{ij}+2\alpha_{,ij}+\alpha_{i,j}^{V}+\alpha_{j,i}^{V}+h^{T}_{ij}]~,
\eea
where $B=\chi-\beta$ and $B^{V}_{i}=\chi^{V}_{i}-\beta^{V}_{i}$. Besides the familiar scalar perturbations ($A, B, \psi, \alpha$), vector perturbations ($B_i^V,
\alpha^V_i$), and tensor perturbations $h^T_{ij}$ in the metric, the parametrization of tetrad brings six extra variables, which are scalar perturbation $\lambda,
\chi+\beta$ and vector perturbation $\lambda_i^V, \chi_i^V+\beta_i^V$. All the vector perturbations are transverse and denoted by the superscript $V$, both the
tensor perturbations are transverse and traceless and denoted by the superscript $T$.
In addition, the scalar field $\theta$ is decomposed as $\theta(\eta, \vec{x})=\bar{\theta}(\eta)+\delta\theta(\eta, \vec{x})$.

Although we can perform a similar decomposition on the Lorentz matrix $\Lambda^{A}_{~B}$ following the parametrization in Ref.~\cite{PVtele2},
we do not need to do so in this paper.
Because we can always transform the perturbed Lorentz matrix
into the background Lorentz matrix in Eq.~(\ref{solution0}) through the infinitesimal Lorentz transformation (\ref{LT}).
In other words, we can always absorb the perturbations of the Lorentz matrix $\Lambda^{A}_{~B}$ into the perturbations of the tetrad $e^{A}_{~\mu}$
through the infinitesimal Lorentz transformation (\ref{LT}), so that we only need to deal with the perturbations of the the tetrad.

Due to the diffeomorphism invariance, it is safe to take the unitary gauge $\delta\theta=0,~\alpha=0,~\alpha_{i}^{V}=0$.
This simplifies the calculations, for example, the gauge invariant scalar perturbation $\zeta=-(\psi+\mathcal{H}\delta\theta/\theta')$ representing the curvature
perturbation of the hypersurfaces of constant $\theta$ reduces to $-\psi$ under the unitary gauge.
Since both $\alpha$ and $\alpha_{i}^{V}$ are perturbations which enter the metric,
the perturbations $\alpha$, $\alpha_{i}^{V}$ and $\delta\theta$ are invariant under the infinitesimal Lorentz transformation (\ref{LT}).
Therefore, the unitary gauge is compatible with
the operation of absorbing the perturbations of the Lorentz matrix into the perturbations of the tetrad.

The non-isotropic nature of the background connection may lead to coupling of scalar, vector and tensor perturbations.
Therefore, when studying linear perturbations around the irregular flat universe (\ref{solution0}),
we should not deal with scalar, vector, or tensor perturbations individually, but should deal with all perturbation variables simultaneously.
In the following we choose $A$, $\zeta$, $B$, $B_{i}^{V}$, $\beta_{i}=\beta_{,i}+\beta_{i}^{V}$,
$\lambda_{i}=\lambda_{,i}+\lambda_{i}^{V}$ and $h^{T}_{ij}$ as independent variables,
and we study the linear perturbations around the irregular flat universe by means of quadratic action.

For the NYTG model (\ref{NYTG}) with $S_m=0$, one can directly obtain the quadratic action as
\bea\label{qaction1}
& &\nonumber
S^{(2)}=\int \zd^{4}x~ a^{2} \bigg\{6\mathcal{H}\zeta'A-3{\zeta'}^{2}-(2A+\zeta)\zeta_{,ii}-a^{2}V A^{2}
+2(\zeta'-\mathcal{H}A)B_{,ii}
+\frac{1}{8}\left(h^{T\prime}_{ij}h^{T\prime}_{ij}-h^{T}_{ij,k}h^{T}_{ij,k}\right)
\\
& &\quad\quad\quad\quad
-\frac{1}{4}B_{i}^{V}B_{i,jj}^{V}
+c\theta'\Big[2\lambda_{i}\zeta_{,i}+\frac{1}{2}\epsilon^{ijk}(\beta_{i}\beta_{j,k}-\lambda_{i}\lambda_{j,k})+\hat{\mathcal{S}}_{ij}\lambda_{i}\beta_{j}
-\frac{1}{2}\epsilon^{ijk}\mathcal{S}_{il}h^{T}_{jl}\beta_{k}-\frac{1}{8}\epsilon^{ijk}h^{T}_{il}h^{T}_{jl,k}\Big]\bigg\}.~~~
\eea
where $\mathcal{S}_{ij}=v_{i}v_{j}f(\eta)F^{(1)}(\vec{v}\cdot\vec{x})$ and
$\hat{\mathcal{S}}_{ij}=(v_{i}v_{j}-v^{2}\delta_{ij})f(\eta)F^{(1)}(\vec{v}\cdot\vec{x})$.
In general, the coefficients $\mathcal{S}_{ij}$ and $\hat{\mathcal{S}}_{ij}$ are explicitly dependent on the spatial coordinate $\vec{x}$ unless $F$ is a linear function of $\vec{v}\cdot\vec{x}$.
It means that generally the evolution equations for the linear perturbations are neither homogeneous nor isotropic.
This is reasonable since we are studying the perturbations around an irregular background.
In this paper we only consider the case where $F(\vec{v}\cdot\vec{x})=\vec{v}\cdot\vec{x}$ for simplicity
\footnote{The expression of $F(\vec{v}\cdot\vec{x})$ can differ by a constant term, which does not change
the coefficients $\mathcal{S}_{ij}$ and $\hat{\mathcal{S}}_{ij}$.
And a constant factor of the difference of $F(\vec{v}\cdot\vec{x})$ can be absorbed into $f(\eta)$.}.
In this way, $\mathcal{S}_{ij}$ and $\hat{\mathcal{S}}_{ij}$ are constant coefficients.
So the evolution equations for the linear perturbations are homogeneous.
We will see later that when all constraints are lifted, the quadratic action (\ref{qaction1}) will become homogeneous and isotropic.
In addition,
the terms $\hat{\mathcal{S}}_{ij}\lambda_{i}\beta_{j}$ and $\epsilon^{ijk}\mathcal{S}_{il}h^{T}_{jl}\beta_{k}$ in
the action (\ref{qaction1}) show that there is a coupling of scalar, vector and tensor perturbations.
But such coupling may be eliminated by the constraints imposed by the action (\ref{qaction1}) itself.
Therefore, only after the constraints are lifted can we know whether there is really a coupling of scalar, vector and tensor perturbations.

To further simplify the quadratic action, we change to the momentum space in terms of Fourier transformations,
\be\label{Ftrans}
\zeta(\eta, \vec{x})=\int \frac{\zd^{3}k}{(2\pi)^{\frac{3}{2}}}\, \zeta(\eta,\vec{k})\,\mathrm{e}^{\mathrm{i}\vec{k}\cdot\vec{x}}~,
\ee
and we also expand the variables $A$, $B$, ${\lambda}_{i}$, ${\beta}_{i}$ and $h^{T}_{ij}$ in the same way.
Note that we use the normal letter $\zi$ for imaginary unit to distinguish it from the italic letter $i$ used for spatial indices.
The tensor perturbation $h^{T}_{ij}$ can be further expanded as
\be\label{Texpand}
h^{T}_{ij}(\eta, \vec{k})=\sum_{\zA}h_{\zA}(\eta, \vec{k})\, \hat{e}_{i j}^{\zA}(\vec{k})~,
\ee
where $\{\hat{e}^{\zA}_{ij}(\vec{k}),~\zA=\zL,\zR\}$ are circular polarization bases
\footnote{
Note that the choice of circular polarization bases is not unique,
$\hat{e}^{\zA}_{ij}(\vec{k})$ can be rotated along the $\vec{k}$-axis while maintaining all the properties of the circular polarization bases.
For the case where there is a constant vector $\vec{v}\neq0$ on the background,
we can always choose the circular polarization bases to satisfy $v_{i}v_{j}\hat{e}^{\zA}_{ij}(\vec{k})=(v^{2}/\sqrt{2})\sin^{2}\vartheta$,
where $\vartheta$ is the angle between $\vec{k}$ and $\vec{v}$.
This choice maximally simplifies the quadratic action (\ref{qaction3}), so we adopt this choice in this paper.
}
satisfying
$\hat{k}^{l} \epsilon_{lik}\hat{e}_{j k}^{\zA}(\vec{k})=\mathrm{i} \pA \hat{e}_{i j}^{\zA}(\vec{k})$,
where $\hat{k}$ is the unit vector of $\vec{k}$, $\vp_{\zL}=-1$ and $\vp_{\zR}=1$.
Note that we use the normal letter $\zA$ for the left- and right- hand indices
to distinguish it from the italic letter $A$ used to represent the tetrad indices.
The quadratic action in the momentum space can be expressed as
\bea\label{qaction2}
& &\nonumber
S^{(2)}=\int \zd\eta\int \zd^{3}k~ a^{2}\bigg\{6\mH\zeta'A^{*}-3\zeta^{*\prime}\zeta'+k^{2}(2A+\zeta)\zeta^{*}+2k^{2}(\mH A-\zeta')B^{*}
\\
& &\nonumber\quad\quad\quad\quad
-a^{2}V A^{*}A
+\frac{1}{4}k^{2}B_{i}^{V*}B_{i}^{V}
+\frac{1}{4}\sum_{\mA}\left[h_{\zA}^{*\prime}h_{\zA}'-(k^{2}-c\theta'\pA k) h^{*}_{\zA}h_{\zA}\right]
\\
& &\quad\quad\quad\quad
+c\theta'\Big[2\zi k^{i}\lambda^{*}_{i}\zeta+\frac{\zi}{2}\epsilon^{ijk}k^{i}(\beta^{*}_{j}\beta_{k}-\lambda^{*}_{j}\lambda_{k})
+\hat{\mathcal{S}}_{ij}\lambda^{*}_{i}\beta_{j}
-\frac{1}{2}\beta^{*}_{i}\Big(\sum_{\zA}\mathcal{S}^{\zA}_{i}h_{\zA}\Big)
\Big]\bigg\},
\eea
where $\mathcal{S}^{\zA}_{i}(\vec{k})=\epsilon_{ijk}{\mathcal{S}}_{jl}\hat{e}^{\zA}_{kl}(\vec{k})$.
It can be seen that $A$, $B$, $B_{i}^{V}$, $\lambda_{i}$ and $\beta_{i}$ are all non-dynamical fields
and the variations of the action (\ref{qaction2}) with them lead to the following constraints:
\bea
& &\label{ceom1} B^{V}_{i}=0~,\\
& &\label{ceom2} \mathcal{H}A-\zeta'=0~,\\
& &\label{ceom3} 3\mathcal{H}\zeta'+k^{2}\zeta-a^{2}VA+\mathcal{H}k^{2}B=0~,\\
& &\label{ceom4} \epsilon_{ijk}k^{j}\lambda_{k}-\mathrm{i}\hat{S}_{ij}\beta_{j}+2k^{i}\zeta=0~,\\
& &\label{ceom5} -\hat{S}_{ij}\lambda_{j}+\mathrm{i}\epsilon_{ijk}k^{j}\beta_{k}+\frac{1}{2}\sum_{A}\mathcal{S}^{\zA}_{i}h_{\zA}=0~.
\eea

For the regular flat universe case with $v_{i}=0$ or $f(\eta)=0$, there are $\hat{S}_{ij}=0$ and $\mathcal{S}^{\zA}_{i}=0$,
so the solution of Eqs. (\ref{ceom1}), (\ref{ceom2}), (\ref{ceom3}), (\ref{ceom4}) and (\ref{ceom5}) is
\be\label{csol0}
\zeta=0~,~A=0~,~B=0~,~B_{i}^{V}=0~,~\lambda_{i}=\mathrm{i}k^{i}\lambda~,~\beta_{i}=\mathrm{i}k^{i}\beta~,
\ee
where $\lambda$ and $\beta$ are arbitrary scalar perturbations.
Substituting the Eq.~(\ref{csol0}) back into the  action (\ref{qaction2}), the action (\ref{qaction2}) can be simplified as
\be\label{qaction0}
S^{(2)}=\int \zd\eta\int \zd^{3}k\,\frac{a^{2}}{4}\sum_{\zA}\left(|h_{\zA}'|^{2}-\omega_{\zA}^{2}|h_{\zA}|^{2}\right)~,
\ee
where $\omega^{2}_{\zA}=k^{2}-c\theta'\pA k$.
It can be seen that there is no scalar dynamical degree of freedom at the linear perturbation level.
This is a bit strange because the action (\ref{NYTG}) clearly shows that there is a scalar dynamical degree of freedom.
Further research in Ref.~\cite{PVtele2} shows that the missing scalar dynamical degree of freedom reappears in the regular curved universe.
The phenomenon of degrees of freedom being hidden under special background
also appears in $f(\mathbb{T})$ gravity \cite{Ong:2013qja} and massive gravity \cite{DeFelice:2012mx}.
This implies that such a special background is likely to suffer from strong coupling issue \cite{Delhom:2022vae}.
It can also be seen that the modified dispersion relation $\omega^{2}_{\zA}$ is helicity dependent.
This means that GWs with different helicities will have different propagation velocities.
This phenomenon is called velocity birefringence, which is a direct reflection of the parity violation in the NYTG model.
These results are consistent with the results in Refs.~\cite{PVtele1,PVtele2}
\footnote{The subtle difference in the dispersion relation $\omega^{2}_{\zA}$
is due to the difference between expanding by $e^{\zi \vec{k}\cdot\vec{x}}$
and expanding by $e^{-\mathrm{i}\vec{k}\cdot\vec{x}}$ in the Fourier transformation.}.

For the irregular flat universe case with $v_{i}\neq0$ and $f(\eta)\neq0$,
the solution of Eqs. (\ref{ceom1}), (\ref{ceom2}), (\ref{ceom3}), (\ref{ceom4}) and (\ref{ceom5}) is
\bea\label{csol1}
& &\nonumber
A=\zeta'/\mathcal{H}~,~B=-\left[\theta^{\prime 2}\zeta'+2k^{2}\mathcal{H}\zeta\right]/2k^{2}\mathcal{H}^{2}~,~B^{V}_{i}=0~,~\\
& &\nonumber
\lambda_{i}=\Big(\frac{2\cos\vartheta}{kv\sin^{2}\vartheta}\epsilon_{ijk}k^{j}v_{k}\Big)\zeta
-\frac{\mathrm{i}}{2\sqrt{2}k}k^{i}\Big(\sum_{\zA}\pA h_{\zA}\Big)~,\\
& &
\beta_{i}=\Big(\frac{2\mathrm{i}}{v^{2}f(\eta)\sin^{2}\vartheta}k^{i}+\frac{2\mathrm{i}vf(\eta)\cos\vartheta}{k\sin^{2}\vartheta}v_{i}\Big)\zeta
+\frac{\mathrm{i}vf(\eta)\cos\vartheta}{2\sqrt{2}k}v_{i}\Big(\sum_{\zA}h_{\zA}\Big)~,
\eea
where $\vartheta$ is the angle between $\vec{k}$ and $\vec{v}$.
Substituting the above results back into the action (\ref{qaction2}), the action (\ref{qaction2}) can be simplified as
\be\label{qaction3}
S^{(2)}=\int \zd\eta\int \zd^{3}k\,\bigg\{
\frac{z^{2}}{2}\left(|\zeta'|^{2}-k^{2}|\zeta|^{2}\right)
+\frac{a^{2}}{4}\sum_{\zA}\left(|h_{\zA}'|^{2}-\omega_{\zA}^{2}|h_{\zA}|^{2}\right)
-\frac{ca^{2}\theta'k}{\sqrt{2}}\zeta^{*}\Big(\sum_{\zA}\pA h_{\zA}\Big)\bigg\}~,
\ee
where $z^{2}=a^{2}\theta'^{2}/\mathcal{H}^{2}$.
For the  action  (\ref{qaction3}), the following points need to be emphasized.
Firstly, it can be seen that there is indeed a scalar dynamical degree of freedom,
which again verifies that there is a scalar dynamical degree of freedom hidden under the regular flat universe at the linear perturbation level.
Secondly, there are two tensor dynamics degrees of freedom and the dispersion relation $\omega_{\zA}^{2}$ is helicity dependent,
as is the case for the regular universe.
This means that the velocity birefringence phenomenon of GWs also exists in the irregular universe.
Thirdly, it is surprising that $v_{i}$ and $f(\eta)$ are completely cancelled in the step of lifting the constraints,
so that the action (\ref{qaction3}) no longer depends on $v_{i}$ and $f(\eta)$.
This makes the case of $v_{i}=0, f(\eta)=0$ not the limit of the case of $v_{i}\rightarrow0, f(\eta)\rightarrow0$.
This is somewhat analogous to the case where a massless photon is not the limit of a photon with mass tends to zero.
Fourth, it can be seen that the coefficients in the action (\ref{qaction3}) are homogeneous and isotropic.
This means that the evolution equations of the scalar perturbation $\zeta$ and the tensor perturbations $h_{\zA}$ are homogeneous and isotropic.
It should be emphasized that such a property depends on the fact that we only consider the simple case with $F(\vec{v}\cdot\vec{x})=\vec{v}\cdot\vec{x}$.
In the next section we will see that this results in isotropic power spectra for perturbations.
This is consistent with observations since the current data from the CMB found no statistically significant evidence for direction dependence of the power spectra \cite{Planck:2018jri}.
More general considerations of the function $F$ and the constrains on it from observational data through detailed analysis deserve further studies.
These considerations are model dependent and beyond the scope of this paper.
Finally, it can be seen that even after the constraints are lifted, there is still a coupling of scalar and tensor degrees of freedom.
This is a feature that neither in the regular flat universe nor in the regular curved universe.
This means that scalar perturbations and tensor perturbations can influence each other at the linear perturbation level.
This can be seen more clearly from the perspective of the equations of motion.
From the action (\ref{qaction3}), the linear equations of $\zeta$ and $h_{\zA}$ can be obtained as
\bea
& &\label{steom1}
\zeta''+2\frac{z'}{z}\zeta'+k^{2}\zeta+\frac{ca^{2}\theta'k}{\sqrt{2}z^{2}}\Big(\sum_{\zA}\pA h_{\zA}\Big)=0~,\\
& &\label{steom2}
h_{\zA}''+2\mathcal{H}h_{\zA}'+\omega_{\zA}^{2}h_{\zA}+\sqrt{2}c\theta'\pA k\zeta=0~.
\eea
Eq.~(\ref{steom1}) shows that the tensor perturbations $h_{\zA}$ can be used as a source of the scalar perturbation $\zeta$.
The scalar perturbation $\zeta$ can be excited when left- and right- handed GWs have different amplitudes or phases.
And Eq.~(\ref{steom2}) shows that the scalar perturbation $\zeta$ can be used as a source of the tensor perturbations $h_{\zA}$.
It is worth noting that the source of the tensor perturbations $h_{\zA}$ caused by $\zeta$ is helicity-dependent,
that is, the excitation effects caused by $\zeta$ on the left- and right-handed GWs are different.

\section{Primordial fluctuations generated by inflation}\label{power spectrum}
In the previous section, we preliminarily studied the the linear perturbations around the regular and irregular flat universe,
and obtained the quadratic action after the constraints was lifted.
In this section, we will  preliminarily study the primordial fluctuations generated by slow-roll inflation in the regular and irregular flat universe.

\subsection{The case of the regular universe}\label{subregular}
For the case of regular universe, the quadratic action (\ref{qaction0}) can be expressed as
\be\label{qaction01}
S^{(2)}=\int \zd\eta\int \zd^{3}k\,\frac{a^{2}}{2}\sum_{\zA}
\left[\Big|\frac{1}{\sqrt{2}}h_{\zA}'\Big|^{2}-\left(k^{2}-c\theta'\pA k\right)\Big|\frac{1}{\sqrt{2}}h_{\zA}\Big|^{2}\right]~.
\ee
Note that since there are only tensor degrees of freedom in the regular flat universe at the linear perturbation level,
a scalar field other than $\theta$ needs to be introduced to generate the primordial scalar perturbation \cite{PVtele1,Cai:2021uup}.
In this subsection we do not consider the case of introducing additional scalar fields, and we only focus on the tensor perturbations.

Next we consider the case of slow-roll inflation dominated by the axion-like field $\theta$.
Since the background equations of the regular flat universe are exactly the same as those in GR,
the background evolution during inflation will be exactly the same as the case of slow-roll inflation in GR \cite{Baumann:2009ds,Wang:2013zva}.
So we don't need to repeat the analysis of the details of single scalar field inflation.
We introduce two commonly used slow-roll parameters
\be\label{slowroll}
\varepsilon\equiv-\frac{\dot{H}}{H^{2}}~,~
\delta\equiv\frac{\ddot{\theta}}{H\dot{\theta}}~,
\ee
where $H=\dot{a}/{a}=\mathcal{H}/a$ is the Hubble rate, the upper dot represents the derivative with respect to the physical time $t$.
We assume $\ce\sim|\delta|\ll1$, $|\dot{\ce}/H|\ll|\ce|$ and $|\dot{\delta}/H|\ll|\delta|$ during inflation.
Under the slow-roll approximation,
\be\label{approx0}
\mathcal{H}\approx -\frac{1+\ce}{\eta}~~,~~\theta'\approx \frac{\sqrt{2\ce}}{\eta}~.
\ee
Without loss of generality, in Eq.~(\ref{approx0}) we have assumed that the value of $\theta$ decreases during inflation.

Next, by combining Eqs~(\ref{qaction01}) and (\ref{approx0}),
the correlation function of $h_{\zA}$ can be obtained through the process in Appendix \ref{Correlation}:
\be\label{correlation01}
\langle h_{\zA}^{\dagger}h_{\zA} \rangle\approx H^{2}\ze^{-\pA\sqrt{\ce/2}c\pi}k^{-(3+2\varepsilon)}~,
\ee
and $\langle h_{\zL}^{\dagger}h_{\zR}\rangle=0$.
Through the correlation functions (\ref{correlation01}), the power spectrum of the left- and right-handed GWs can be obtained as
\be\label{PA0}
\mathcal{P}_{\zA}(k)=\frac{k^{3}}{\pi^{2}}\langle h_{\zA}^{\dagger}h_{\zA}\rangle\approx
\frac{H^{2}}{\pi^{2}}\ze^{-\pA\sqrt{\ce/2}c\pi}k^{-2\varepsilon}~.
\ee
The power spectrum of the tensor perturbations can be obtained as
\be\label{Ptensor}
\mathcal{P}_{T}(k)=\mathcal{P}_{\zL}(k)+\mathcal{P}_{\zR}(k)\approx\frac{H^{2}}{\pi^{2}}\left[1+\cosh\left(\sqrt{\frac{\ce}{2}}c\pi\right)\right]k^{-2\ce}~.
\ee
The relative different between the power spectrum of the left- and right-handed GWs can be obtained as
\be\label{reladiff0}
\Pi\equiv\frac{\mathcal{P}_{\zR}-\mathcal{P}_{\zL}}{\mathcal{P}_{\zR}+\mathcal{P}_{\zL}}\approx
-\tanh\left(\sqrt{\frac{\ce}{2}}c\pi\right)\approx -\sqrt{\frac{\ce}{2}}c\pi~.
\ee
$\Pi\neq0$ means that the magnitudes of the primordial fluctuations of left- and right-handed GWs are different.
This is a clear physical signal of parity violation.
But this seems to contradict the conclusion in Refs.~\cite{PVtele1,PVtele2} that
there is only velocity birefringence of GWs but no amplitude birefringence of GWs in the NYTG model.
The reason for this contradiction is that $\theta'$ is approximated as a constant in the analysis of the evolution of GWs in Refs.~\cite{PVtele1,PVtele2}.
Of course, this approximation is valid when studying the propagation of GWs in a slowly expanding universe.
However, $\theta'=a\dot{\theta}\propto 1/\eta$ cannot be approximated as a constant during the slow-roll inflation dominated by $\theta$.
We know that for a harmonic oscillator (the equation of motion is $\ddot{x}+\omega^{2}x=0$),
the amplitude of the harmonic oscillator can be changed when the frequency $\omega$ is time-dependent.
And when the time dependence of $\theta'$ is not negligible,
the time dependence of $\omega_{L}$ and $\omega_{R}$ will be different,
resulting in different effects on the amplitudes of left- and right-hand GWs.
This is why the magnitudes of the primordial fluctuations
of left- and right-handed GWs generated by slow-roll inflation in the regular flat universe are different.
If $\ce\rightarrow0$, it can be seen from Eq.~(\ref{approx0}) that $\theta'\approx0$ can be approximated as a constant,
and from Eq.~(\ref{reladiff0}), it can be seen that $\Pi\rightarrow0$ too, that is,
the magnitudes of the primordial fluctuation of the left- and right-handed GWs are the same.

Finally, let's look at the case when the coupling constant $c\rightarrow0$, then
\be\label{GRcase0}
\mathcal{P}_{T}(k)\approx \frac{2H^{2}}{\pi^{2}}k^{-2\ce}~~,~~\Pi\approx0~.
\ee
This is exactly the result of the slow-roll inflation of single scalar field in GR.

\subsection{The case of the irregular universe}\label{subirregular}

For the case of irregular universe,
since the coupling of $\zeta$ and $h_{\zA}$ in the action (\ref{qaction3})
makes it difficult to analyze the quantum fluctuations, we first diagonalize the variables $\zeta$ and $h_{\zA}$ below.
Firstly, for the convenience of analysis,
we introduce new variables $\xi_{1}=(z/a)\zeta$, $\xi_{2}=(1/\sqrt{2})h_{\zL}$ and $\xi_{3}=(1/\sqrt{2})h_{\zR}$,
so that the action (\ref{qaction3}) can be simplified as
\be\label{qaction4}
S^{(2)}=\int \zd\eta\int \zd^{3}k\,\frac{a^{2}}{2} \bigg\{
\sum_{\zs=1}^{3}\xi^{*\prime}_{\zs}\xi_{\zs}-\sum_{\zs_{1}=1}^{3}\sum_{\zs_{2}=1}^{3}\mathbf{M}_{\zs_{1}\zs_{2}}\xi^{*}_{\zs_{1}}\xi_{\zs_{2}}
\bigg\}~,~~\text{with}~~
\mathbf{M}=
\left(
  \begin{array}{ccc}
    k^{2}-\Omega & -\kappa & \kappa \\
    -\kappa & k^{2}-\sigma & 0 \\
    \kappa & 0 & k^{2}+\sigma \\
  \end{array}
\right)~,
\ee
where $\Omega=z''/z-a''/a$, $\sigma=-c\theta'k$ and $\kappa=c\mathcal{H}k$ are background quantities.
Secondly, we introduce an orthogonal matrix $\mathbf{T}$ that can diagonalize the matrix $\mathbf{M}$, and its expression is
\be\label{orthogonalmatrix}
\mathbf{T}=\left(
             \begin{array}{c}
               \mathbf{t}_{1}^{\zT} \\
               \mathbf{t}_{2}^{\zT} \\
               \mathbf{t}_{3}^{\zT} \\
             \end{array}
           \right)~,~\text{with}~~
\mathbf{t}_{\zs}=
\frac{-\zs^{2}+5\zs-5}
{\sqrt{1+\frac{(\tau_{\zs}-\sigma)^{2}}{\kappa^{2}}+\left(1-\frac{(\tau_{\zs}-\sigma)(\tau_{\zs}+\Omega)}{\kappa^{2}}\right)^{2}}}
               \left(
                 \begin{array}{c}
                   (\tau_{\zs}-\sigma)/\kappa \\
                   1-(\tau_{\zs}-\sigma)(\tau_{\zs}+\Omega)/\kappa^{2} \\
                   1 \\
                 \end{array}
               \right)~,
\ee
where the superscript $\zT$ means transpose, and $\{\tau_{\zs},~\zs=1,2,3\}$ are the solutions of the cubic equation
\be\label{eigenEq}
\tau^{3}+\Omega\tau^{2}-(2\kappa^{2}+\sigma^{2})\tau-\sigma^{2}\Omega=0~.
\ee
The specific expressions of $\{\tau_{\zs},~\zs=1,2,3\}$ are in Appendix \ref{eigensolution}.
Finally, we introduce new variables $\{q_{\zs},~\zs=1,2,3\}$, which are defined as
\be\label{newvariable}
\left(
  \begin{array}{c}
    q_{1} \\ q_{2} \\ q_{3} \\
  \end{array}
\right)
= \mathbf{T}
\left(
  \begin{array}{c}
    \xi_{1} \\ \xi_{2} \\ \xi_{3} \\
  \end{array}
\right).
\ee
Thus, the action (\ref{qaction4}) can be further simplified as
\be\label{qaction5}
S^{(2)}=\sum_{\zs=1}^{3} \int \zd\eta\int \zd^{3}k\,\frac{a^{2}}{2}
\Big\{|q_{\zs}'|^{2}-(k^{2}+\tau_{\zs})|q_{\zs}|^{2}\Big\}~.
\ee
So far, we have simplified the action (\ref{qaction3}) with coupling between variables to the action (\ref{qaction5}) without coupling between variables.
The latter form makes it easier to calculate the primordial fluctuations generated by inflation.

Next we consider the case of slow-roll inflation dominated by the axion-like field $\theta$.
Since in Sec.~\ref{Nonregular flat universe}
we proved that the background equations of the irregular flat universe are exactly the same as those in GR,
the background evolution during inflation will be exactly the same as the case of slow-roll inflation in GR.
Under the slow-roll approximation, the background quantities $\Omega$, $\sigma$ and $\kappa$ can be approximately expressed as
\be\label{approxB}
\Omega\approx \frac{3(\ce+\delta)}{2\eta^{2}}~,~\sigma\approx -\frac{\sqrt{2\ce}ck}{\eta}~,~\kappa\approx-\frac{(1+\ce)ck}{\eta}~.
\ee
In this section, we also assume that the coupling constant $c\sim1$  (it can also be seen as a requirement of naturalness),
so that $c\gg\sqrt{\ce}$.
Ignoring high-order small quantities such as $\ce^{2}$, $\{\tau_{\zs},~\zs=1,2,3\}$ in Eq.~\reff{apEigenvalues} can be approximated as
\be\label{approxtau}
\tau_{1}\approx \frac{(2+3\ce)ck}{\sqrt{2}\eta}-\frac{3(\ce+\delta)}{2\eta^{2}}~,~
\tau_{2}\approx 0~,~
\tau_{3}\approx -\frac{(2+3\ce)ck}{\sqrt{2}\eta}-\frac{3(\ce+\delta)}{2\eta^{2}}~.
\ee
If only up to the order of $\sqrt{\ce}$ is retained, the orthogonal matrix $\mathbf{T}$ can be approximated as
\be\label{approxT}
\mathbf{T}\approx
\left(
  \begin{array}{ccc}
    \frac{1}{\sqrt{2}} & \frac{1+\sqrt{\ce}}{2} & -\frac{1-\sqrt{\ce}}{2} \\
    -\sqrt{\ce} & \frac{1}{\sqrt{2}} & \frac{1}{\sqrt{2}} \\
    \frac{1}{\sqrt{2}} &  -\frac{1-\sqrt{\ce}}{2} & \frac{1+\sqrt{\ce}}{2} \\
  \end{array}
\right)
\ee
Regarding the approximate expression (\ref{approxT}),
there are two points that need additional explanation.
First, the order $\sqrt{\ce}$ is the lowest order approximation required to preserve the difference in the power spectrum of left- and right-handed GWs.
If we further ignore the contribution of $\sqrt{\ce}$ in $\mathbf{T}$,
the difference in the power spectrum of left- and right-handed GWs disappears.
And if we keep the higher-order terms, it brings only more complex but less important corrections in the power spectrum.
Second, it can be seen that the matrix $\mathbf{T}$ does not tend to the identity matrix as $c\rightarrow0$ in the approximate expression (\ref{approxT}).
This is confusing because the three variables are all decoupled as $c\rightarrow0$ in the action (\ref{qaction4}).
The reason for this confusing phenomenon is that we have used the approximation $c\gg\sqrt{\ce}$ in Eqs.~(\ref{approxtau}) and (\ref{approxT}).
If $c$ is too small, neither the Eq.~(\ref{approxtau}) nor Eq.~(\ref{approxT}) hold.
See Appendix \ref{ctendsto0} for the approximate behavior of orthogonal matrix $\mathbf{T}$ when $c\rightarrow0$.

Next, by combining Eqs~(\ref{qaction5}) and (\ref{approxtau}),
the correlation function between variables $q_{\zs}$ can be obtained through the process in Appendix \ref{Correlation}:
\be\label{correlation1}
\langle q_{1}^{\dagger}q_{1}\rangle\approx\frac{H^{2}}{2}\ze^{\frac{c\pi}{\sqrt{2}}}k^{-(3+3\varepsilon+\delta)}~,~
\langle q_{2}^{\dagger}q_{2}\rangle\approx\frac{H^{2}}{2}k^{-(3+2\varepsilon)}~,~
\langle q_{3}^{\dagger}q_{3}\rangle\approx\frac{H^{2}}{2}\ze^{-\frac{c\pi}{\sqrt{2}}}k^{-(3+3\varepsilon+\delta)}~,
\ee
and $\langle q_{\zs_{1}}^{\dagger}q_{\zs_{2}}\rangle=0$ when $\zs_{1}\neq\zs_{2}$.
Then, using the approximation techniques in Appendix \ref{Aptechnique} and combining Eqs~(\ref{newvariable}), (\ref{approxT}) and (\ref{correlation1}),
the correlation functions for the variables $\zeta$ and $h_{\zA}$ can be obtained as
\bea\label{correlation2}
& &\nonumber
\langle \zeta^{\dagger}\zeta\rangle\approx \frac{1}{2\ce}\cosh\left(\frac{c\pi}{\sqrt{2}}\right) H^{2}k^{n_{S}-4}~,
\\
& &\nonumber
\langle h_{\zA}^{\dagger}h_{\zA}\rangle\approx
\left[\frac{1}{2}+\frac{1}{2}\cosh\left(\frac{c\pi}{\sqrt{2}}\right)-\pA\sqrt{\ce}\sinh\left(\frac{c\pi}{\sqrt{2}}\right)\right] H^{2}k^{n_{T}-3}~,
\\
& &\nonumber
\langle \zeta^{\dagger}h_{\zA}\rangle\approx
-\frac{\pA}{2\sqrt{2\ce}}\sinh\left(\frac{c\pi}{\sqrt{2}}\right)H^{2}k^{-(3+3\ce+\delta)}~,
\\
& &
\langle h_{\zL}^{\dagger}h_{\zR}\rangle\approx \frac{1}{2}\left[1-\cosh\left(\frac{c\pi}{\sqrt{2}}\right)\right]H^{2}
k^{-(3+3\ce+\delta)-\frac{1}{2}\,\csch^{2}\left(\frac{c\pi}{2\sqrt{2}}\right)(\ce+\delta)}~,
\eea
where
\be\label{STindex}
n_{S}\approx1-(\delta+3\ce)~~,~~
n_{T}\approx-(3\ce+\delta)+\frac{1}{2}\,\sech^{2}\left(\frac{c\pi}{2\sqrt{2}}\right)(\ce+\delta)~.
\ee
It should be noted that since Eqs.~(\ref{approxtau}) and (\ref{approxT}) are approximately true only when $c\gg\sqrt{\ce}$,
Eqs.~(\ref{correlation2}) and (\ref{STindex}) are also approximately true only when $c\gg\sqrt{\ce}$.

Through the correlation functions (\ref{correlation2}),
the power spectrum of the scalar perturbation $\zeta$ can be obtained as
\be\label{Pscalar}
\mathcal{P}_{S}(k)=\frac{k^{3}}{2\pi^{2}}\langle \zeta^{\dagger}\zeta\rangle \approx
\frac{H^{2}}{8\pi^{2}\ce}\cosh\left(\frac{c\pi}{\sqrt{2}}\right)k^{n_{S}-1}~.
\ee
The power spectrum of the left- and right-handed GWs can be obtained as
\be\label{PA}
\mathcal{P}_{\zA}(k)=\frac{k^{3}}{\pi^{2}}\langle h_{\zA}^{\dagger}h_{\zA}\rangle\approx
\frac{H^{2}}{2\pi^{2}}\left[1+\cosh\left(\frac{c\pi}{\sqrt{2}}\right)-2\pA\sqrt{\ce}\sinh\left(\frac{c\pi}{\sqrt{2}}\right)\right]k^{n_{T}}.
\ee
The power spectrum of the tensor perturbations can be obtained as
\be\label{Ptensor}
\mathcal{P}_{T}(k)=\mathcal{P}_{\zL}(k)+\mathcal{P}_{\zR}(k)\approx\frac{H^{2}}{\pi^{2}}\left[1+\cosh\left(\frac{c\pi}{\sqrt{2}}\right)\right]k^{n_{T}}~.
\ee
The tensor-to-scalar ratio $r$ can be obtained as
\be\label{TSratio}
r\equiv\frac{\mathcal{P}_{T}}{\mathcal{P}_{S}}=8\left[1+\sech\left(\frac{c\pi}{\sqrt{2}}\right)\right]\ce~.
\ee
The relative different between the power spectrum of the left- and right-handed GWs can be obtained as
\be\label{reladiff}
\Pi\equiv\frac{\mathcal{P}_{\zR}-\mathcal{P}_{\zL}}{\mathcal{P}_{\zR}+\mathcal{P}_{\zL}}\approx-2\sqrt{\ce}\tanh\left(\frac{c\pi}{\sqrt{2}}\right)~.
\ee
Strictly speaking, since Eqs.~(\ref{correlation2}) and (\ref{STindex}) are only approximately true when $c\gg\sqrt{\ce}$,
Eqs.~(\ref{Pscalar})-(\ref{reladiff}) are also approximately true only when $c\gg\sqrt{\ce}$.
But If we ignore this fact and force $c\rightarrow0$, then
\be\label{GRcase}
\mathcal{P}_{S}\approx \frac{H^{2}}{8\pi^{2}\ce}k^{n_{S}-1}~,~
\mathcal{P}_{T}\approx \frac{2H^{2}}{\pi^{2}}k^{n_{T}}~,~
r\approx 16\ce~,~
\Pi\approx 0~.
\ee
It can be seen that except for the spectral indices $n_{S}$ and $n_{T}$, Eq.~(\ref{GRcase}) is the result of the slow-roll inflation in GR.

\begin{figure}[h]
  \centering
  \includegraphics[width=0.47\textwidth]{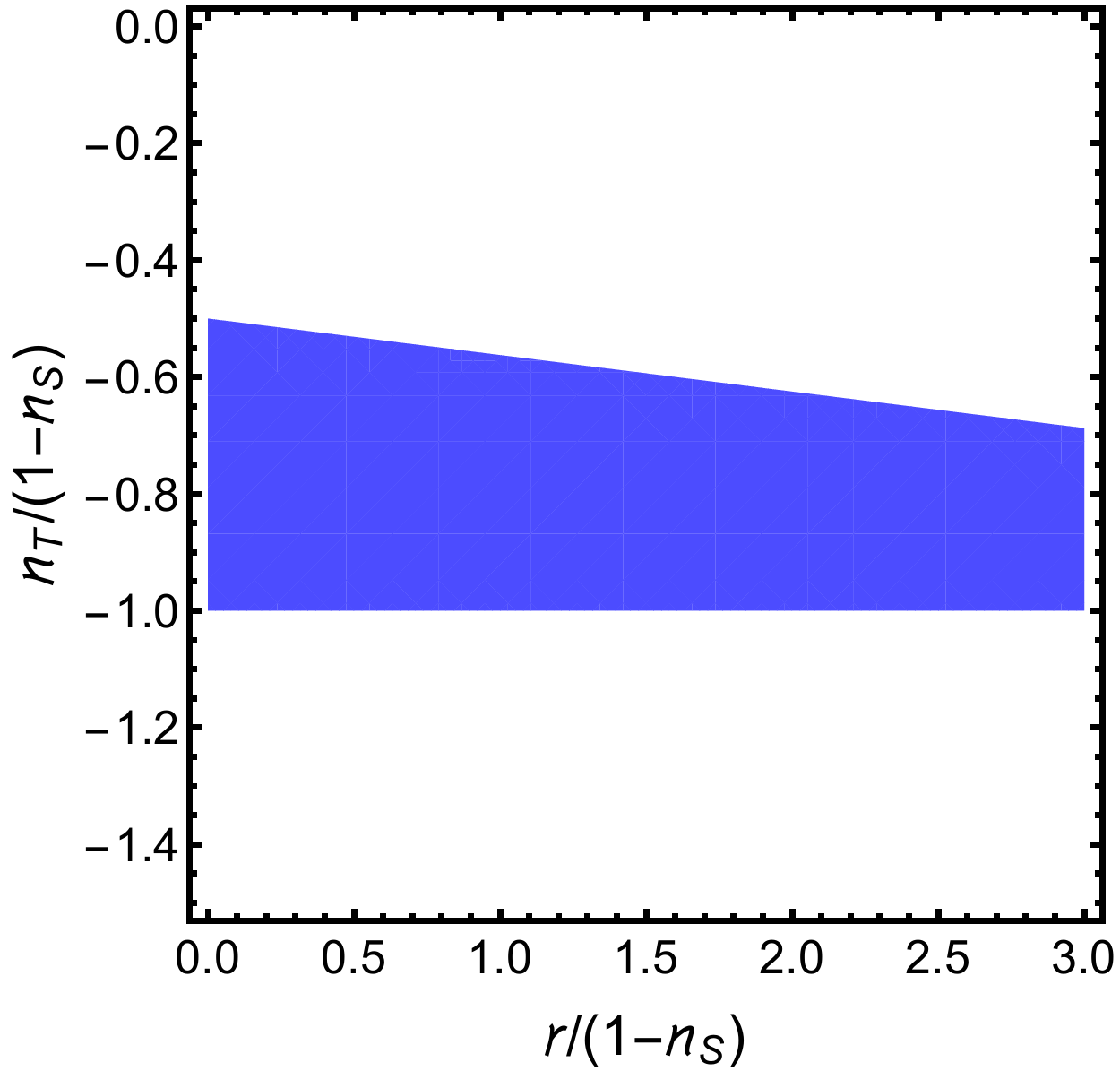}\\
  \caption{
  The blue region is the region where the coupling constant $c$ is real.
  Since \emph{Planck} 2018 \cite{Planck:2018nkj} gives $r<0.101$ and $n_{S}\approx 0.966$, we only considered the interval $0<r/(1-n_{S})<3$.
  }
  \label{figure1}
\end{figure}

Once the spectral index $n_{S}, n_{T}$ and the tensor-to-scalar ratio $r$ are measured experimentally,
we can obtain the slow-roll parameters $\ce, \delta$ and the coupling constant $c$ as
\bea
& &
\ce=\frac{1}{16}\Big[4(1-n_{S})+r+\sqrt{(r+4-4n_{S})^{2}+16r\, n_{T}}\Big]~,
\\ & &
\delta=\frac{1}{16}\Big[4(1-n_{S})-3r+3\sqrt{(r+4-4n_{S})^{2}+16r\, n_{T}}\Big]~,
\\ & &\label{csol}
c=\frac{\sqrt{2}}{\pi}\arccosh\bigg[\frac{\sqrt{(r+4-4n_{S})^{2}+16r\, n_{T}}-4(1-n_{S})-8n_{T}}{8(1-n_{S}+n_{T})}\bigg]~.
\eea
In order to ensure that $c$ given by Eq.~(\ref{csol}) is real,
we need the values of $n_{S}, n_{T}$ and $r$ to fall in the blue region in FIG.~\ref{figure1}.
If the values of $n_{S}, n_{T}$ and $r$ fall outside the blue region in FIG.~\ref{figure1}, it means that $c$ given by Eq.~(\ref{csol}) is not real,
and this irregular universe model cannot be consistent with the experiment.
If the values of the $n_{S}, n_{T}$ and $r$ fall in the blue region in FIG.~\ref{figure1},
then we can obtain the values of $\ce, \delta$ and $c$,
so that we can calculate the value of the relative different $\Pi$ between the left- and right-handed primordial GWs  through Eq.~(\ref{reladiff}).
On this basis, if the relative different $\Pi$ can be measured experimentally,
we can test whether this irregular universe model is consistent with the experiment by comparing the theoretical and experimental values of $\Pi$.

\begin{figure}[h]
  \centering
  \includegraphics[width=0.5\textwidth]{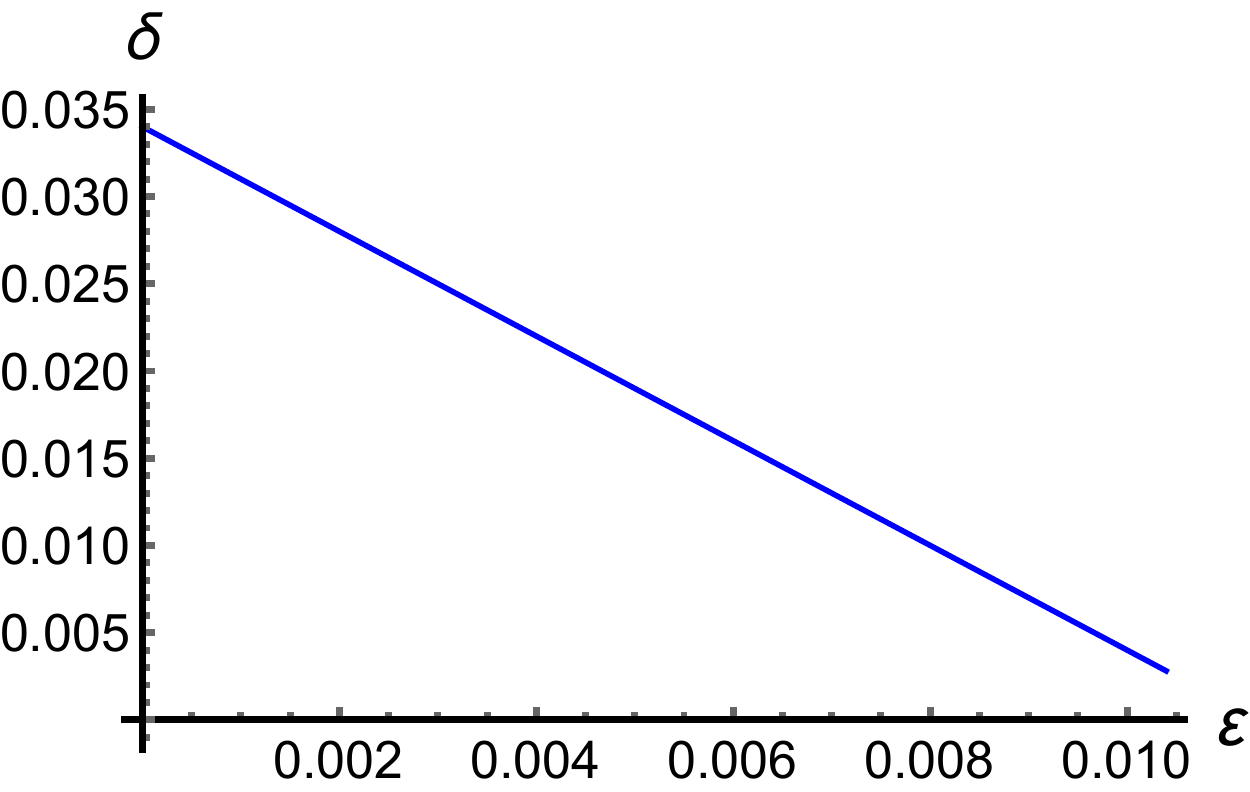}\\
  \caption{In the $\ce$-$\delta$ plane, the blue line is the allowable value range when $c=1$.}
  \label{figure2}
\end{figure}

Unfortunately, primordial GWs have not yet been observed experimentally,
so we cannot know the values of tensor spectral index $n_{T}$, the tensor-to-scalar ratio $r$ and
the relative different $\Pi$ between the left- and right-handed primordial GWs.
From the \emph{Planck} 2018 \cite{Planck:2018nkj},
we only know that the scalar spectral index $n_{S}\approx 0.966$ and the tensor-to-scalar ratio $r<0.101$.
Therefore, all we can know is that the allowable value range of the slow-roll parameters $\ce,\delta$ is
\be
0<\ce<\frac{0.101}{8\left[1+\sech\left(c\pi/\sqrt{2}\right)\right]}<0.012625~~,~~
\delta\approx0.034-3\ce~.
\ee
The maximum value of $\ce$ depends on the coupling constant $c$, but will not exceed $0.012625$
(the upper limit of $\ce$ when $c\rightarrow\infty$).
The allowable value of $\delta$ is determined by $\ce$.
FIG.~\ref{figure2} shows the allowable value range of slow-roll parameters $\ce,\delta$ when $c=1$.

Although by comparing the results in subsections \ref{subregular} and \ref{subirregular},
we can find that the power spectrum of the left- and right-handed GWs given by the irregular universe is different from that of the regular universe.
But this is not the main difference between irregular and regular universes for primordial fluctuations.
For primordial fluctuations, the most important feature of the irregular universe compared to the regular universe is that
the correlation function of scalar perturbation and tensor perturbations
$\langle \zeta^{\dagger}h_{\zA}\rangle\neq0$ at the linear perturbation level.
This means that there is a strong statistical correlation between primordial scalar fluctuations and primordial tensor fluctuations
generated by slow-roll inflation in the irregular universe.
The apparent reason for this phenomenon is that the quadratic action contains the coupling of scalar perturbations and tensor perturbations
in the irregular universe, as exhibited by the action (\ref{qaction3}).
The deeper reason may be that the condition $\mathcal{L}_{\xi_I}\hat{\Gamma}^{\rho}_{~\mu\nu}\neq0$
destroys the homogeneity and isotropy of the interior space, so that the scalar fluctuations and the tensor fluctuations
can interact with each other in the irregular universe.

\section{Conclusion}\label{conclusion}

As a step towards exploring the irregular universe within the TG framework, in this paper, we studied the irregular flat universe of the NYTG model.
Firstly, we obtained the irregular flat universe solution of the NYTG model under the condition that only the symmetry of the metric is required.
We found that the cosmological background equations of the NYTG model are exactly the same as those of GR
in both the regular flat universe and the irregular flat universe.
Secondly, we studied the linear cosmological perturbations around the irregular flat universes.
We found a peculiar feature of the irregular flat universe: the tensor and scalar perturbations are coupled together at the linear perturbation level.
We speculate that this peculiar feature is caused by the fact that
the interior space does not satisfy the homogeneity and isotropy in the irregular universe.
Finally, we applied the NYTG model to the early universe and studied the primordial perturbations generated by slow-roll inflation in the regular and irregular flat universes.
We found that the left- and right-handed primordial GWs are different in both the regular flat universe and the irregular flat universe.
We also found that there is a strong statistical correlation between the primordial scalar and tensor perturbations
generated by slow-roll inflation in the case of irregular universe, this is a direct consequence of the direct coupling between the scalar and tensor perturbations at linear order.

{ \it Acknowledgement}:
This work is supported in part by National Key R\&D Program of China Grant No. 2021YFC2203102, and by NSFC under Grant No. 12075231 and 12047502.

\appendix

\section{Solutions of the cubic equation}\label{eigensolution}
Consider a cubic equation with respect to the variable $\tau$ as
\be\label{apEq1}
\va\tau^{3}+\vb\tau^{2}+\vc\tau+\vd=0~,
\ee
where $\va$, $\vb$, $\vc$ and $\vd$ are real coefficients.
In order to express the solution of Eq.~(\ref{apEq1}) conveniently, we introduce the following parameters
\be
\vA=\vb^{2}-3\va\vc~,~\vB=\vb\vc-9\va\vd~,~\vC=\vc^{2}-3\vb\vd~,~\Delta=\vB^{2}-4\vA\vC~,~
\Theta=\frac{1}{3}\arccos\left(\frac{2\vA\vb-3\vB\va}{2\vA^{3/2}}\right)~.
\ee
When $\Delta<0$, Eq.~(\ref{apEq1}) has three real solutions, which are
\bea\label{apEigenvalues}
& &\nonumber
\tau_{1}=-\frac{1}{3\va}\left(\vb+2\sqrt{\vA}\cos\Theta\right)~,
\\
& &\nonumber
\tau_{2}=\frac{1}{3\va}\left[-\vb+\sqrt{\vA}\left(\cos\Theta-\sqrt{3}\sin\Theta\right)\right]~,
\\
& &
\tau_{3}=\frac{1}{3\va}\left[-\vb+\sqrt{\vA}\left(\cos\Theta+\sqrt{3}\sin\Theta\right)\right]~,
\eea

The Eq.~(\ref{eigenEq}) in the main text is the result of taking $\va=1$, $\vb=\Omega$, $\vc=-(2\kappa^{2}+\sigma^{2})$
and $\vd=-\sigma^{2}\Omega$ in Eq.~(\ref{apEq1}).
In this case, there are always $\vA\geq0$ and $\Delta\leq0$,
where the equal sign holds if and only if $\kappa=\sigma=\Omega=0$.
And when $\kappa=\sigma=\Omega=0$, obviously the three solutions of Eq.~(\ref{eigenEq}) are $\tau_{1}=\tau_{2}=\tau_{3}=0$,
and the orthogonal matrix $\mathbf{T}$ is the identity matrix.

\section{The orthogonal matrix $\mathbf{T}$ when $c\rightarrow0$}\label{ctendsto0}

In this appendix, we discuss the approximate behavior of the orthogonal matrix $\mathbf{T}$ in Eq.~(\ref{orthogonalmatrix})
as $c\rightarrow0$ in a more general background (not only during inflation).
Since $\sigma\propto c$ and $\kappa\propto c$, then
\be\label{apapproxB}
\frac{\kappa}{\Omega}\propto c~,~\frac{\sigma}{\Omega}\propto c~,~\frac{\kappa^{2}}{2\sigma\Omega}\propto c~.
\ee
When $c$ is much smaller than any other background quantities such as $\sqrt{\ce}$, $\dot{\theta}$ and $\mathcal{H}^{-1}$,
ignoring the quadratic and higher terms of $c$, the solutions of Eq.~(\ref{eigenEq}) can be approximately expressed as
\be\label{apapproxtau}
\tau_{1}\approx\Omega~,~\tau_{2}\approx\sigma~,~\tau_{3}\approx\sigma~.
\ee
So the orthogonal matrix $\mathbf{T}$ in Eq.~(\ref{orthogonalmatrix}) can be approximately expressed as
\be\label{apapproxT}
\mathbf{T}=
\left(
  \begin{array}{ccc}
    1 & \frac{\kappa}{\Omega} & -\frac{\kappa}{\Omega}\\
    -\frac{\kappa}{\Omega} & 1 & \frac{\kappa^{2}}{2\sigma\Omega} \\
    \frac{\kappa}{\Omega} & -\frac{\kappa^{2}}{2\sigma\Omega} & 1 \\
  \end{array}
\right)
\xrightarrow{\text{when}~c\rightarrow0}
\left(
  \begin{array}{ccc}
    1 & 0 & 0 \\
    0 & 1 & 0 \\
    0 & 0 & 1 \\
  \end{array}
\right)
\ee
It can be easily seen from Eqs.~(\ref{apapproxB}) and (\ref{apapproxT}) that
when $c\rightarrow0$, the orthogonal matrix $\mathbf{T}$ does tend to the identity matrix.
This is consistent with the fact that all variables in the action (\ref{qaction4}) tend to be decoupled when $c\rightarrow0$.

\section{Correlation function generated by inflation}\label{Correlation}
The purpose of this appendix is to show how to calculate the correlation function generated by inflation.
Consider a univariate system whose effective action during inflation is
\be\label{apaction1}
S=\frac{1}{2} \int \zd\eta\,\zd^{3}k\, a^{2} \left[ |q_{\vec{k}}'|^{2}-\left(k^{2}-\frac{2\va k}{\eta}-\frac{3\vb}{\eta^{2}}\right)|q_{\vec{k}}|^{2}\right]~,
\ee
where $\va$ and $\vb$ are real parameters, and $\vb$ has the same order of magnitude as the slow-roll parameter $\varepsilon$.
Here $q(\eta,\vec{x})$ is the variable and we have changed to the  Fourier space $q_{\vec{k}}(\eta)$.
After quantization, the variable $q_{\vec{k}}(\eta)$ can be expanded as
\be\label{apexpand}
q_{\vec{k}}(\eta)=\frac{1}{a(\eta)}\left(v_{k}(\eta)\hat{a}_{\vec{k}}+v_{k}^{*}(\eta)\hat{a}_{\vec{k}}^{\dagger}\right)~,
\ee
where $\hat{a}_{\vec{k}}^{\dagger}$ and $\hat{a}_{\vec{k}}$ are the generation and annihilation operators
that satisfy the following commutation relations
\be\label{apcommutation}
[\hat{a}_{\vec{k}}~ \hat{a}_{\vec{k}'}^{\dagger}]=\updelta^{(3)}(\vec{k}-\vec{k}')~~,~~
[\hat{a}_{\vec{k}}~ \hat{a}_{\vec{k}'}]=[\hat{a}_{\vec{k}}^{\dagger}~ \hat{a}_{\vec{k}'}^{\dagger}]=0~,
\ee
and $v_{k}(\eta)$ satisfies the following equation
\be\label{apeom1}
v_{k}''+\left(k^{2}-\frac{2\va k}{\eta}-\frac{\mu^{2}-1/4}{\eta^{2}}\right)v_{k}=0~,
\ee
where $\mu\approx 3/2+\varepsilon+\vb$.
Note that in Eq.~(\ref{apeom1}), we used the approximation $a''/a\approx [(3/2+\varepsilon)^{2}-1/4]/\eta$,
and we ignored the higher-order terms of $\varepsilon$ and $\vb$.
Next we choose the Bunch-Davies vacuum at $\eta\rightarrow-\infty$, that is,
\be
\lim_{\eta\rightarrow-\infty}v_{k}=\frac{1}{\sqrt{2k}}\mathrm{e}^{-\mathrm{i}k\eta}~.
\ee
Under this condition, the solution for Eq.~(\ref{apeom1}) is (for more detail, see \cite{Satoh:2010ep})
\be
v_{k}(\eta)=\ze^{-\mathrm{i}k\eta}(-2k\eta)^{\mu}(-\eta)^{\frac{1}{2}}\ze^{-\mathrm{i}\pi(\frac{1}{4}+\frac{\mu}{2})}
\mathbf{U}\left(1/2+\mu-\zi\va, 1+2\mu;~ 2\zi k\eta\right)\ze^{-\frac{\va\pi}{2}}~,
\ee
where $\mathbf{U}(c_{1},c_{2};\, \mathrm{z})$ is the confluent hypergeometric function.
The $|v_{k}|$ has the following asymptotic form when $k\eta\rightarrow 0^{-}$ (super-horizon scale)
\be\label{apasymptotic}
|v_{k}|\approx 2^{\mu-1}\pi^{-\frac{1}{2}}\Gamma(\mu)k^{-\mu}(-\eta)^{\frac{1}{2}-\mu}\ze^{-\frac{\va\pi}{2}}
\approx 2^{-\frac{1}{2}}\ze^{-\frac{\va\pi}{2}}aHk^{-\mu}
\ee
where $\Gamma(\mathrm{z})$ is the Gamma function.
In the last approximately equal sign in Eq.~(\ref{apasymptotic}), we used the approximations $\mu\approx3/2$ and $(-\eta)^{-1}\approx aH$.
Combining Eqs.~(\ref{apexpand}), (\ref{apcommutation}) and (\ref{apasymptotic}),
we can obtain the correlation function on the super-horizon scale as
\be\label{correlation0}
\langle0|q^{\dagger}_{\vec{k}}q_{\vec{k}'}|0\rangle\approx
\frac{H^{2}}{2}\ze^{-\va\pi}k^{-(3+2\varepsilon+2\vb)}\updelta^{(3)}(\vec{k}+\vec{k}')~.
\ee
where $|0\rangle$ is the vacuum state, which satisfies $\hat{a}_{\vec{k}}|0\rangle=0$.
For the sake of convenience, we can omit the subscript $\vec{k}$ and throw away the annoying delta function $\updelta^{(3)}(\vec{k}+\vec{k}')$,
so that the correlation function (\ref{correlation0}) can be abbreviated as
\be\label{correlation}
\langle q^{\dagger}q \rangle\approx\frac{H^{2}}{2}\ze^{-\va\pi}k^{-(3+2\varepsilon+2\vb)}~.
\ee

\section{Summation of nearly scale-invariant functions}\label{Aptechnique}
Consider there are $N$ nearly scale-invariant functions $\{\zf_{\vi}(k)=C_{\vi}k^{n_{\vi}},~\vi=1,2,...,N\}$, where $|n_{\vi}|\ll1$.
Then the sum of these functions should also be a nearly scale-invariant function, so it can be approximated as
\be\label{approxtech1}
\zf(k)=\sum_{\vi=1}^{N} \zf_{\vi}(k) = \sum_{\vi=1}^{N} C_{\vi} k^{n_{\vi}} \approx Ck^{n}~,~\text{with}~|n|\ll1~.
\ee
Next we need to find the coefficient $C$ and the exponent $n$ in Eq.~(\ref{approxtech1}).
Since $n_{\vi}\approx0$ and $n\approx0$,  we can approximately let $n_{\vi}=n=0$ in Eq.~(\ref{approxtech1}), so that Eq.~(\ref{approxtech1}) becomes
\be\label{approxtechC}
C\approx \sum_{\vi=1}^{N}C_{\vi}~.
\ee
Next, let Eq.~(\ref{approxtech1}) take the derivative of $k$ and then let $n_{\vi}=n=0$ on the exponent of $k$.
Then the approximate expression of $n$ can be obtained as
\be\label{approxtechn}
n\approx \frac{1}{C}\sum_{\vi=1}^{N}C_{\vi}n_{\vi}~.
\ee

{}


\end{document}